\documentclass[aps,reprint,amsmath,amssymb,superscriptaddress,nofootinbib]{revtex4-2}

\usepackage{graphicx}
\usepackage{amsmath}
\usepackage{gensymb}
\usepackage{upgreek}
\usepackage{xcolor}
\usepackage{tabularx}
\usepackage{array}
\usepackage{physics}
\usepackage{siunitx}
\AtBeginDocument{\RenewCommandCopy\qty\SI}
\usepackage{hyperref}
\usepackage{subcaption}
\usepackage[T1]{fontenc}
\usepackage{cleveref}
\usepackage{enumitem}
\usepackage{xcolor}
\usepackage{comment}

\graphicspath{{./figures/APSD/}}

\begin{document}

\title{Bayesian and geometric analyses of power spectral densities of spin qubits in Si/SiGe quantum dot devices}

\author{Yujun Choi$^{\dagger}$}
\affiliation{Department of Physics, Virginia Tech, Blacksburg, VA 24061, United States}
\affiliation{Virginia Tech Center for Quantum Information Science and Engineering, Blacksburg, VA 24061, United States}

\author{Hruday Mallubhotla$^\dagger$}
\affiliation{Department of Physics, University of Wisconsin-Madison, Madison, WI 53706, United States}

\author{Mark Friesen}
\affiliation{Department of Physics, University of Wisconsin-Madison, Madison, WI 53706, United States}

\author{Susan N.\ Coppersmith}
\affiliation{School of Physics, The University of New South Wales, Sydney, NSW 2052, Australia}

\author{Robert Joynt}\email{rjjoynt@wisc.edu}
\affiliation{Department of Physics, University of Wisconsin-Madison, Madison, WI 53706, United States}

\def\thefootnote{$\dagger$}\footnotetext{These authors contributed equally to this work}

\date{\today} 

\begin{abstract}
Charge noise with a power-law spectrum 
poses a significant challenge to high-fidelity operation of spin qubits in semiconductor devices. Recently, considerable experimental work characterized this noise using qubits as spectrometers. 
It apparently arises from a collection of two-level fluctuating electric dipole systems (TLS). This suggests using the data to infer the positions, orientations, and other physical characteristics of the TLS. We identify a fundamental difficulty in this program: the inference of the TLS parameters is strongly undetermined, since the quantity of data is not sufficient to fix them uniquely.  We describe two approaches to deal with this situation. The first approach is a qualitative method based on analytic calculations and simulations of small model systems that recognizes certain patterns.  The second approach, more appropriate for detailed data analysis, is a Bayesian computation that assigns probabilities to candidate dipole configurations. We propose that the Brier score, a measure of confidence in the probabilities, can be used as a quantitative tool to judge the efficacy of experimental noise-measurement setups.  Together, the analytical and computational Bayesian methods constrain, but do not fix, the density, the positions, the orientations, and the strengths of the dipole noise sources.
\end{abstract}

\maketitle


\section{Introduction}
Charge noise has attracted a great deal of interest due to its adverse effect on semiconductor spin qubit devices \cite{burkard2023semiconductor}.  It shortens decoherence times of the spin qubits and leads to  worse gate fidelities. This degrades the performance of quantum information processing. The noise power typically follows an approximate power-law shape ($1/f^\alpha$) over a wide range of frequencies \cite{struck2020low, connors2022charge, rojas2024origins}. Very often there are deviations from this in the form of some dependence of $\alpha$ on frequency or modifications of the power-law behavior by slightly stronger or weaker noise in certain frequency ranges.  The origin of the noise remains under active investigation, but two-level systems (TLS) that produce a fluctuating electric dipole near the qubits are generally thought to dominate this type of noise \cite{muller2019towards}. In this paper we will assume that this is the case.  

Qubit spectroscopy gives information about the noise.  This comes in the form of the auto power spectral density (APSD) of single qubits and the cross power spectral density (CPSD) of two qubits. The information contained in these quantities is all that is presently available to understand the physical nature of the TLS.  

Some questions about the TLS for which one would like answers are:
\begin{itemize}[leftmargin=0pt]
\item[]
\textit{ Total number.} Compared to the original Dutta-Horn model, in which a large number of TLS follow a uniform distribution of activation energies, recent experiments observed that relatively few TLS can produce the noise spectra observed in experiments \cite{connors2019low, ye2024characterization}.
We need to consider a wide range of total numbers of TLS.
\item[]
\textit{Orientation.}
The direction of the dipole associated with the TLS will also affect the APSD \cite{choi2022anisotropy}.
For example, if the TLS are traps that empty and fill to a nearby  2-dimensional electron gas (2DEG), as hypothesized in Ref.~\cite{sakamoto1995distributions}, the orientation will be predominantly perpendicular to the 2DEG layer.  This trap picture has been around for many decades and is generally associated with the name of McWhorter \cite{mcwhorter1957semiconductor}. TLS due to crystal defects that create two near-equivalent potential minima (possibly, but not necessarily, in the oxide layer)  are also a strong possibility. This is the defect model of Anderson, Halperin, Varma, and Phillips \cite{ anderson1972anomalous,phillips1972tunneling}. Here a more random distribution of orientations is a natural supposition \citep{campbell2002density, hofheinz2006individual}.
These arguments indicate that dipole orientation, which is not often emphasized in theoretical analyses, supplies a very important cue as to the origin of TLS. 
\item[]
\textit{Location.}
The actual spatial position of the dipoles is clearly a central question \cite{yoneda2023noise, rojas2023spatial}. It bears on their physical nature, but proximity to the qubits also determines the relative importance of a TLS since the noise strength falls off rapidly with distance, as will be discussed further below. 
Some likely candidates for the location (or locations) of the TLS are oxide layers, semiconductor-oxide interfaces \cite{Neyens2024-lj}, and metal-semiconductor interfaces. 
Even the Si/SiGe interface is an intriguing possibility:  the chemical similarity of Si and SiGe would seem to make such defects unlikely, but close proximity to the qubit means that even a few such defects could have an outsize importance.
\item[]
\textit{Strength.}
There has also been a considerable body of experimental work that aims to determine the magnitude of the dipole moment of the TLS \cite{shehata2023modeling, kkepa2023simulation, kkepa2023correlations}.  Again, this determination might help to distinguish between the trap and defect models, since the electron would be expected to move further in the trap model, with a correspondingly larger dipole moment.  
\item[]
\textit{Relaxation times.} The determination of relaxation times proceeds by the investigation of the frequency and temperature dependence of the APSD and CPSD.  The temperature ($T$) and frequency ($f$) dependence of the APSD $S(f,T)$ in the Dutta-Horn model is $S\propto T/f$.  This is very often close to what is observed \cite{freeman2016linear}, and almost always close enough that the best way to analyze the data is in terms of deviations from this form.  An interesting recent example of this kind is in Ref.~\cite{connors2019low}. To explain the frequency dependence, several theoretical models have been suggested \cite{ahn2021microscopic, mickelsen2023effects, throckmorton2024generalized}.  One can fit the data, but a complete explanation is still lacking.  
\end{itemize}

We contend that for a solid understanding of the TLS, all of these parameters: number, orientation, location, strength, and relaxation times are necessary.  Limiting a model by fixing one or several of them at the start is risky, and we will attempt to avoid this as far as possible.

Most of the work on charge noise has concerned the APSD of the qubits \cite{yoneda2018quantum, struck2020low, kranz2020exploiting, connors2022charge, jock2022silicon, spence2023probing}, and this has given rise to a rich phenomenology. The CPSD contains information on the spatial correlation of the noise.  It is crucial to study CPSD because many quantum error correcting codes rely on the assumption that there are no correlated errors \cite{devitt2013quantum,brun2019quantum}.  Correlated noise can induce correlations in the errors of multiple qubits.  This effectively reduces the code distance.  Understanding the noise correlations therefore impacts the design of the devices, placing requirements on the inter-dot spacing, for example.  For the purposes of this paper, however, the importance of the CPSD is that it can throw light on the nature of the TLS, as will be seen below.

The CPSD has been measured in dedicated experiments designed partially for this purpose \cite{connors2022charge}, and in qubit experiments \cite{boter2020spatial}.  These studies found weak to moderate values of the CPSD.  More recent work found stronger CPSD, even when the qubits were spaced up to 100 nm and above \cite{yoneda2023noise, rojas2023spatial, donnelly2024noise}.  In some cases specific TLS configurations were found that could reproduce the CPSD 
 and the APSD, usually working with models in which the TLS do not interact.  More elaborate models include interactions \cite{mickelsen2023interacting, zou2024spatially}. We do not analyze interacting models in this paper, but we will point out some aspects of certain experimental results that  suggest they may be present. 

 Just this brief survey references a great deal of data and many intriguing theoretical ideas. Yet overall, the history of the subject still makes a complex and even somewhat confusing picture. This is no doubt partly due to the fact that the TLS are of several kinds and that there is a great deal of sample dependence.  In
 this regard, the discovery that only a relatively small number of TLS is needed to produce the observed noise becomes very significant.  Small numbers are associated with non-universal behavior since they lead to a lack of statistical self-averaging.  This in turn makes it difficult to discern persistent patterns in the data.

We contend in this paper that there is another obstacle in the long road to understanding charge noise in semiconductor quantum dot devices. This is simply that the mathematical problem of deducing the configuration of the TLS from the APSD and the CPSD is strongly underdetermined.  By ``configuration'' we mean a value for the above list of basic parameters: number, orientation, location, strength, and relaxation times.

Faced with this difficult situation one must resort to qualitative or probabilistic methods.  It is not presently feasible to generate a hypothesis about the TLS and use the data to unambiguously confirm or falsify it.  One can only assign probabilities to a hypothesis.  This is still useful, since it gives a guide for future work that could ultimately change probabilities into certainties.  This sort of progress will require a continual conversation between experiment and theory.  

The outline of this work is as follows.  In Sec.~\ref{sec:model} we delineate the universe of possibilities for the configurations, the standard first step in any probabilistic theory.  We also make the connection between the TLS configurations and the qubit spectroscopy observations. In Sec.~\ref{sec:under} we quantify the extent to which the problem of determining dipole configurations is underdetermined by present experimental data. In Sec.~\ref{sec:analytical} qualitative considerations that develop intuition about dipole configurations are given.  Some analytic calculations in the continuum limit are performed, which give some qualitative guidance as to how to narrow down the range of possibilities for the configurations.  We also do some small-scale simulations on artificial data to illustrate how the CPSD can throw light on the geometry of the noise sources. In particular, one can get some information about dipole orientations. Here we focus more on total number and dipole magnitude.
Sec.~\ref{sec:bayes}
establishes the Bayesian formalism that yields configuration probabilities from experimental data and applies it to some specific recent experiments.  Finally, Sec.~\ref{sec:conclusion} recaps our results and indicates some future directions. 

\section{Model of Noise Sources}
\label{sec:model}

\subsection{Parameterization}
The noise model considered in this paper is a set of non-interacting TLS.  The $n$th TLS has a position $\mathbf{r}_n =(x_n,y_n,z_n)$ and a  relaxation time $\tau_n$. Each TLS is an electric dipole flipping back and forth along a fixed orientation. Thus the model is a set of parameters for the TLS, not a completely definite physical picture.  The trap and defect pictures mentioned above are specific examples of such pictures.  Still, if these parameters are fixed one can completely compute the noise they produce.  Success in determining them would be a step in the direction of understanding their physical nature. 
 
We define the growth direction of the devices as the $z$ axis and the quantum dot qubits are in the $z=0$ plane.  In the formulas of this section there are two qubits located at 
$\mathbf{R}_{1} =(x_1,y_1,0)$ and $\mathbf{R}_{2} =(x_2,y_2,0)$.

We begin with a completely general mathematical model in order to stress that the methods used are not limited to any particular device geometry.  We will narrow this down later as needed. 

\subsection{Probability Distribution}

The TLS are distributed according to a probability distribution $\rho(x,y,z,p,\hat{p},\tau)$.  The $n$th TLS flips randomly between states $s_n = \pm 1$ resulting in a time-dependent dipole $\mathbf{p}_n = p \hat{p}_n s_n(t)$.  
The magnitude $p =|\mathbf{p}|$ is taken as a constant in this section for simplicity but we will also vary it below.  

 The methods we use are easily generalized to other dipole models.
 
 We will consider various probability distributions for the orientations $\hat{p}_n$, as will be explained further below.
The definition of $\rho$ is general enough to include correlations between the spatial and temporal parameters $\mathbf{r}$ and $\tau$. Recent data \cite{ye2024characterization} may suggest the existence of such correlations.

\subsection{Noise Correlation Functions}
\label{subsec:ncf}
The quantities of greatest direct interest for spin qubit decoherence are the voltage correlations and the electric field correlations. The conversion of the voltage or field to qubit operating frequency varies from one platform to another. A discussion may be found in Ref.~\cite{choi2024interacting}.  
Our methods apply with minor changes also to magnetic noise, assuming the noise sources are fluctuating magnetic dipoles.

The noisy voltages on the two qubits are given by  
\begin{align}
	V_{1,2}(t) & = 
 -p \sum_n 
 \frac{\hat{p}_n \cdot
 (\mathbf{r}_n - \mathbf{R}_{1,2})}
 {4 \pi \epsilon|(\mathbf{r}_n - \mathbf{R}_{1,2})|^3}
 s_n(t)~.
 \label{eq:voltage}
\end{align}
Here $\epsilon$ is the dielectric constant of the material.  This expression is easily generalized to include image dipoles (screening effects).

It is instructive to think of Eq.~\ref{eq:voltage} in terms of a random walk. The left-hand side is the measured quantity. It is the end point of a walk that is a sum of random vectors, since $\hat{p}_n$ and $\mathbf{r}_n$ are random quantities.  Reconstructing 
the TLS configuration from the measurement is the exercise of reconstructing every step in the walk from only a knowledge of the endpoints. This is already a hint that the problem is underdetermined. 

The noise correlations are  
\begin{align}
	\langle V_1(t)  V_2(0) \rangle & = 
 \left ( \frac{p}{4 \pi \epsilon} \right )^2
 \,
 \sum_{m,n} A_{mn} \,
 \langle s_m(t)  s_n(0) \rangle~.
 \label{eq:apsd}
\end{align}
Angle brackets denote a thermal and quantum average. $A_{mn}$ is a matrix that encodes the geometric information for the electric noise in the system.  It is given by
\begin{align}
	A_{mn}(t) & = 
 \frac{\hat{p}_m \cdot
 (\mathbf{r}_m - \mathbf{R}_1) \,
 \hat{p}_n \cdot
 (\mathbf{r}_n - \mathbf{R}_2)}
 {|(\mathbf{r}_m - \mathbf{R}_1)|^3
 |(\mathbf{r}_n - \mathbf{R}_2)|^3}~.
 \label{eq:amn}
\end{align}
This expression is valid even if the TLS are not statistically independent. 
In this paper we consider only the special case of  non-interacting and therefore statistically independent TLS:
$\langle s_m(t) s_n(0) \rangle = \delta_{mn} \langle s_n(t) s_n(0) \rangle$ so that
\begin{align}
	\langle V_1(t)  V_2(0) \rangle & = 
  \left ( \frac{p}{4 \pi \epsilon} \right )^2
 \,
 \sum_{n} A_{nn} \,
 \langle s_n(t)  s_n(0) \rangle,
 \label{eq:ind}
\end{align}
so that only the diagonal elements of $A$ are important. If in addition all the TLS have the same relaxation time, then
\begin{align}
	\langle V_1(t)  V_2(0) \rangle & = 
 \left ( \frac{p}{4 \pi \epsilon} \right )^2
 \,
 \sum_{n} A_{nn} \,
 \exp(-|t|/\tau).
 \label{eq:tau}
\end{align}

We will consider several different distributions below for $\hat{p}$, the directions of the dipoles \cite{muller2019towards}, but in this section we will only look at two cases.  The first is a fixed direction, say the $x$-direction, in which case $\hat{p}_{n,i} \hat{p}_{m,j}$ is replaced by $\delta_{ix} \delta_{jx}$.  The second is random directions, in which case $\hat{p}_{n,i} \hat{p}_{m,j}$ is replaced by $\delta_{ij}/3$.  Here $i$ and $j$ are Cartesian indices.

To obtain field-field correlations, we can simply differentiate the formulas for the voltage-voltage correlations:
The noise correlations are  
\begin{align}
	\langle E_{1,i}(t)  E_{2,j}(0) \rangle & = 
 \left ( \frac{p}{4 \pi \epsilon} \right )^2
 \,
 \sum_{m,n} B_{mn} \,
 \langle s_m(t)  s_n(0) \rangle~,
 \label{eq:ffcorr}
\end{align}
where $B_{mn} = \partial^2 A_{mn} / \partial \mathbf{R}_1 \partial \mathbf{R}_2$. 
The gradients are taken with respect to the $i$th and $j$th components of the position of qubits 1 and 2 respectively.

The unnormalized CPSD is defined in the frequency domain:
\begin{eqnarray}
	S_{12}(f) &=& \langle V_1(t)  V_2(0) \rangle_{f} \hfil \nonumber \\
    & = &
 \sum_{m,n} A_{mn} \,
 \int dt e^{2 \pi i f t} \langle s_m(t)  s_n(0) \rangle~.
 \label{eq:omega}
\end{eqnarray}

The APSD $S_1(f)$ is simply the limit when $\mathbf{R}_1 = \mathbf{R}_2 $ in Eq.~\ref{eq:omega}.

The normalized CPSD is defined by 
\begin{equation}
\label{eq:CPSD}
C_{12} (f) = \frac{S_{12}(f)}
{\sqrt{S_1 (f) S_2 (f)}}.   
\end{equation} 
This is the standard Pearson formula and it conforms to the notation in 
Refs.~\cite{yoneda2023noise, rojas2023spatial}.  
$C_{ij} (f)$ will be referred to simply as the CPSD from now on.  

The CPSD is a complex-valued function in general \cite{szankowski2017environmental}, giving rise to $C_{12} (f) = c(f) e^{i\gamma (f)}$ where $c(f)$ is the correlation strength and $\gamma(f)$ is the phase of the CPSD.  However, in our model of the TLS $C_{12}$ is a real-valued function when the TLS are statistically independent, as follows immediately from Eq.~\ref{eq:tau}.  In this case $\gamma(f) = 0$ or $\gamma(f) = \pi$.

\section{Underdetermination}
\label{sec:under}

\subsection{Introduction}
To this point, we have focused on formalism: how to define probability distributions for the TLS configurations and what these distributions imply for the observed noise. 

As stated in the introduction, the models in our analysis involve four types of parameters: the orientation of the dipoles (which can be horizontal, vertical, or randomly orientated), the number of dipoles, the magnitude of the dipole moments, and their relaxation times.
Horizontal orientation dipoles are given a random orientation within the plane. A configuration is defined as a situation in which all of these parameters are fixed. 

We would now like to move to the detailed consideration of actual experiments and what they tell us about the TLS. The ideal is to determine the configuration given a set of noise data.  A little reflection reveals that this hope is unrealistic.  Let us fix the number of TLS $N_T$.  For a real device $N_T > 6$.  Then counting the number of parameters needed to specify a configuration tells us that we need to search a space of dimension greater than 42.  This is \textit{before} we vary $N_T$.  The points in this space that correspond to the data, which itself has error bars, live in a submanifold of very high dimension as well. We may characterize the problem as like trying to find a low-dimensional needle in a high-dimensional haystack.

\subsection{Example}
 In this section we give a concrete example of the pitfalls that exist when the problem of underdetermination is ignored. We do this by trying to solve a mock problem with known TLS parameters.
 
 A tempting way to fit experimental data is to make some plausible assumptions to construct a parameterized model, and then to use a deterministic optimization approach to get values for these parameters that optimize the fit of the model to the data. 

The difficulty with this very standard way of proceeding is that good fits to the data can be obtained that may in fact be divorced from physical reality. The assumptions may have seemed innocent, but when they are incorrect, the calculation will not detect it.  In the presence of underdetermination, a model will fix a manifold in the parameter space that is large enough to fit the data even though that manifold does not contain the actual solution.  This is true even for quite simple cases, as we now show.

We use a dimensionless cost function to compare the PSD $S_c$ of a TLS configuration to a measured PSD $S_m$:
\begin{equation}
\label{eq:cost}
	C(S_c(f)) = \frac{1}{\log f_u- \log f_{\ell}} \int^{f_u}_{f_{\ell}} \frac{df}{f} \frac{\left(S_m(f) - {S_c}(f)\right)^2}{\sigma(f)^2}.
\end{equation}
This form for $C$ finds the cost in terms of an average number of standard deviations $\sigma$. It weights frequency decades equally, which is the natural procedure if the distribution of TLS relaxation rates is log-uniform. We choose the ratio of maximum $f_u$ to minimum $f_{\ell}$ frequencies to be $ f_u/f_{\ell} = 10^4$.

We take the artificial data to be a single TLS with given position, orientation, and magnitude. It has a position $\left(x, y, z = h \right)$ where $h$ is assumed to be known, as well as a dipole moment $\vec{p} = \left(p_x, p_y, p_z\right)$. It is oriented in the $z$-direction.  These are the parameters to be found. 

We compute its exact PSD which is then measured at two separate dots, giving us our test data to be fit.  We model the TLS in different ways to be described below and then perform a least-squares analysis to see if we can determine the parameters of the TLS.

For a candidate set of parameters, the cost is computed between the measured PSD and the candidate's generated noise spectrum, and solutions are obtained by finding the values which minimize this cost.

\begin{figure*}[ht]
\centering
\setkeys{Gin}{width=\linewidth}
	\begin{subcaptionblock}[b]{.48\textwidth}
		\includegraphics[width=\linewidth]{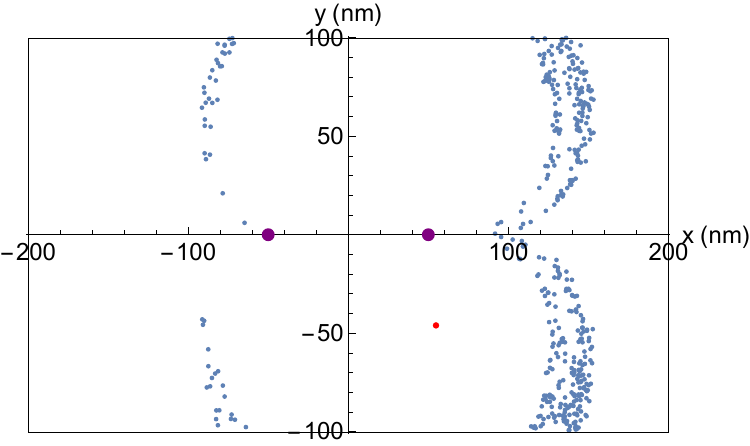}
		\caption{
			Incorrect orientation assumption.
		}
		\label{fig:ch4:singledipolematch_twodot_0.1_0}
	\end{subcaptionblock}%
	\begin{subcaptionblock}[b]{.48\textwidth}
		\includegraphics[width=\linewidth]{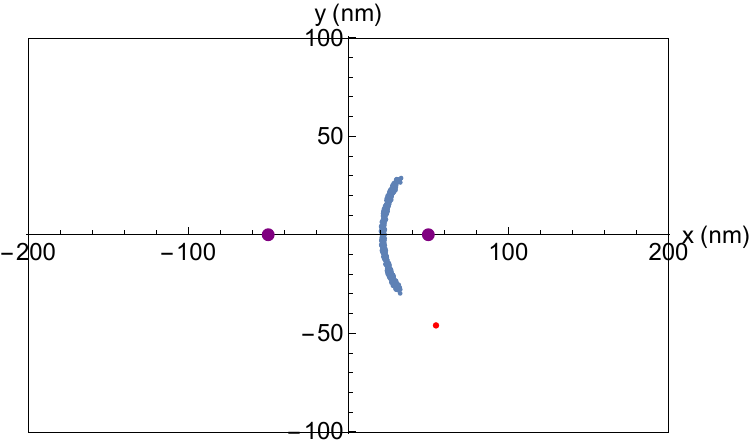}
		\caption{
			Incorrect magnitude assumption.
		}
		\label{fig:ch4:singledipolematch_twodot_0.1_4}
	\end{subcaptionblock}%
	\caption{
		Effect of incorrect parameter assumptions on TLS position estimation.
		These figures demonstrate how incorrect assumptions about dipole orientation (a) or magnitude (b) still lead to solutions that match the measured APSD data.
		The blue points in each panel denote configurations where the total cost function defined in Eq. \ref{eq:cost}, summed over the APSDs measured at the two dots, is smaller than 0.1. In (a), TLS positions assuming horizontal dipoles are calculated and plotted in blue, while the correct dipole (red) is oriented in the $z$-direction, normal to the plane.
		In (b), the dipole moment used to predict TLS positions is actually one tenth of the actual TLS (red). The calculated solutions are in blue.
		In both cases, a simple cost-minimization approach is giving low-cost solutions for the TLS configuration that do not resemble the actual configuration. 
	}
	\label{fig:ch4:singledipolematch_incorrect_parameters}
\end{figure*}

In 
Fig.~\ref{fig:ch4:singledipolematch_incorrect_parameters}
the large solid purple circles are qubits (with $z=0$) that measure the noise, the small red circle is the TLS (with $z=h$), the blue dots represent positions of the TLS that are consistent (have a cost below a certain threshold) with the experimental data under a given model.  

Our first analysis assumes incorrectly that the TLS is oriented in the $x-y$ plane.
This leaves three parameters to be found, the dipole magnitude $p$ and horizontal position $x, y$.
The result is shown in Fig. ~\ref{fig:ch4:singledipolematch_incorrect_parameters} (a).  In spite of a completely wrong assumption, the data can still be fit, as shown by the presence of many solutions with low cost.
The problem is worsened by the possibility of rotation of the dipole in the plane.

Our second analysis contains the incorrect assumption that the magnitude of the dipole moment is 1/10 of the actual value. 
 Fig.\ref{fig:ch4:singledipolematch_incorrect_parameters} (b) shows that many low-cost solutions are obtained with this grossly incorrect value for the magnitude.  The points in blue are solutions $x, y$ that have the incorrect choice of $p$. Hence naive fitting gives the wrong result for the position and the magnitude of the TLS.

If the magnitude $p$ is chosen incorrectly, ideally no solutions should be possible. In this case, however, solutions with the incorrect magnitude are found.

This shows that the fundamental limitation of attempting to solve an underdetermined problem is present even for a \textit{single} TLS.
A real device with dozens of TLSs would prove even less tractable, not to mention varying the number of TLSs, taking into account the possibility of non-uniform TLS magnitudes and so on.

We have increased the number of qubits in this sort of approach and do find that the number of incorrect solutions decreases.  However, really meaningful results begin to appear only when there are about 6 qubits, a huge number to investigate a single TLS.

\section{Analytical Approaches}
\label{sec:analytical}

Our conclusion so far is that the inverse problem of deducing the TLS configurations from the observations is in practice undetermined.  Nevertheless, we can hope for some qualitative or semi-quantitative information. In this section we focus on what can be obtained from analytical analysis and small-scale simulations.  

\subsection{Continuum Limit}
\label{subsec:continuum}
\subsubsection{Introduction}
\label{subsubsec:intro}
The first case we treat is when the number of TLSs is very large. Then it is useful to think of the probability distribution $\rho(x,y,z,\hat{p},\tau)$ as a continuous function and evaluate the PSDs by integration of Eq.~\ref{eq:apsd}.  This gives insight into the geometric aspects of the noise correlations. We will also discuss when the continuum picture breaks down, which turns out to shed light on why it seems that in many devices relatively few TLSs dominate the noise.

We note that of all the arguments of the function $\rho$, $\tau$ is special: each data point in the APSD is associated with a certain range of $\tau$, which is not true of the other variables. Thus the geometry of the noise is best studied by fixing a definite $\tau$, as we will do in this section.  This gives a contribution to the PSD in the range $f \leq 1/\tau$.  For computational purposes, one notes simply that the PSD is additive, so one can simply bin the $\tau$-values and add the results, (which are Lorentzians if the decay is purely exponential as in Eq.~\ref{eq:tau}) obtained for the various bins.  In addition we will take the TLS to be independent.  Modifying Eq.~\ref{eq:tau} to incorporate these assumptions  gives 
\begin{align}
\label{eq:vv}
	\langle V_1(t)  V_2(0) \rangle & = 
 \left ( \frac{p}{4 \pi \epsilon} \right )^2
 \,
 \exp(-|t|/\tau) \,
 A (\mathbf{R}_1,\mathbf{R}_2).
 \end{align}
In this equation the continuum
limit of the matrix $A_{mn}$ has been used.  This is the averaged version $A$ with only two arguments after it is averaged over the probability distribution $\rho$:
 \begin{align}
  A (\mathbf{R}_1,\mathbf{R}_2)
  =\int d^3\mathbf{r}
  \int d \hat{p} \,
 \rho(\mathbf{r},\hat{p}) A(\mathbf{r}_m=\mathbf{r}_n,\hat{p}, \mathbf{R}_1,\mathbf{R}_2).
 \label{eq:gdef}
\end{align}
The electric field correlation tensor under the same assumptions is 
\begin{align}
\label{eq:ee}
	\langle \mathbf{E}_{1}(t) \mathbf{E}_2 (0) \rangle & = 
 \left ( \frac{p}{4 \pi \epsilon} \right )^2
 \,
 \exp(-|t|/\tau) \,
 \nabla_{\mathbf{R_1}}
 \nabla_{\mathbf{R_2}}
 A (\mathbf{R}_1,\mathbf{R}_2).
 \end{align}

\subsubsection{Integral Expressions}
\label{subsubsec:integral}
For the concrete examples in this paper we focus on models where the TLSs are located at or near a gate interface in an oxide layer. In the former case the $N$ dipoles are assumed to inhabit a layer of infinitesimal thickness. Thus they are all at a height $h$ and are uniformly distributed in a circle $\mathcal{C}$ of radius $R$ with coordinates $0 \leq \sqrt{x^2+y^2} \leq R $, so that $\rho(x,y,z)= N \delta(z-h)/ (\pi R^2) $ if  $0 \leq \sqrt{x^2+y^2} \leq R $ and $\rho = 0$ otherwise. We take the qubits to be at $ \mathbf{R}_1 = (-d/2,0,0)$ and $ \mathbf{R}_2 = (d/2,0,0)$. Their midpoint is at the origin, directly beneath the center of the circle of TLS.  

Using Eqs.~\ref{eq:vv} and \ref{eq:gdef}, the geometric parts of the voltage correlation functions are given by 
\begin{align}
\label{eq:g1}
A_x(\mathbf{R}_1,\mathbf{R}_2) & = 
  \int_{\mathcal{C}} dx \, dy \, 
  \frac{x^2-d^2/4}{D(d,x,y)} \\
  \label{eq:g2}
  A_y(\mathbf{R}_1,\mathbf{R}_2) &  = 
  \int_{\mathcal{C}} dx \, dy \,
  \frac{y^2}{D(d,x,y)} \\
  \label{eq:g3}
 A_z(\mathbf{R}_1,\mathbf{R}_2) & = 
 \int_{\mathcal{C}} dx \, dy \,  
  \frac{h^2}{D(d,x,y)} \\
   A_r(\mathbf{R}_1,\mathbf{R}_2) &  = 
\int_{\mathcal{C}} dx \, dy \,  
  \frac{x^2-d^2/4+y^2+h^2}{3D(d,x,y)}. 
\label{eq:g4}
\end{align}
Here the subscripts on $x,y,z$ on $A$ indicate the corresponding Cartesian directions for the dipole orientations while $r$ indicates random orientation.
We note that $A_r=(A_x+A_y+A_z)/3$. 
The denominator is 
\begin{equation}
\label{eq:denominator}
D(d,x,y) = 
[(x+d/2)^2+y^2+h^2]^{3/2}
[(x-d/2)^2+y^2+h^2]^{3/2}.
\end{equation}

\begin{figure}[ht]
\centering
\includegraphics[width=0.48\textwidth]{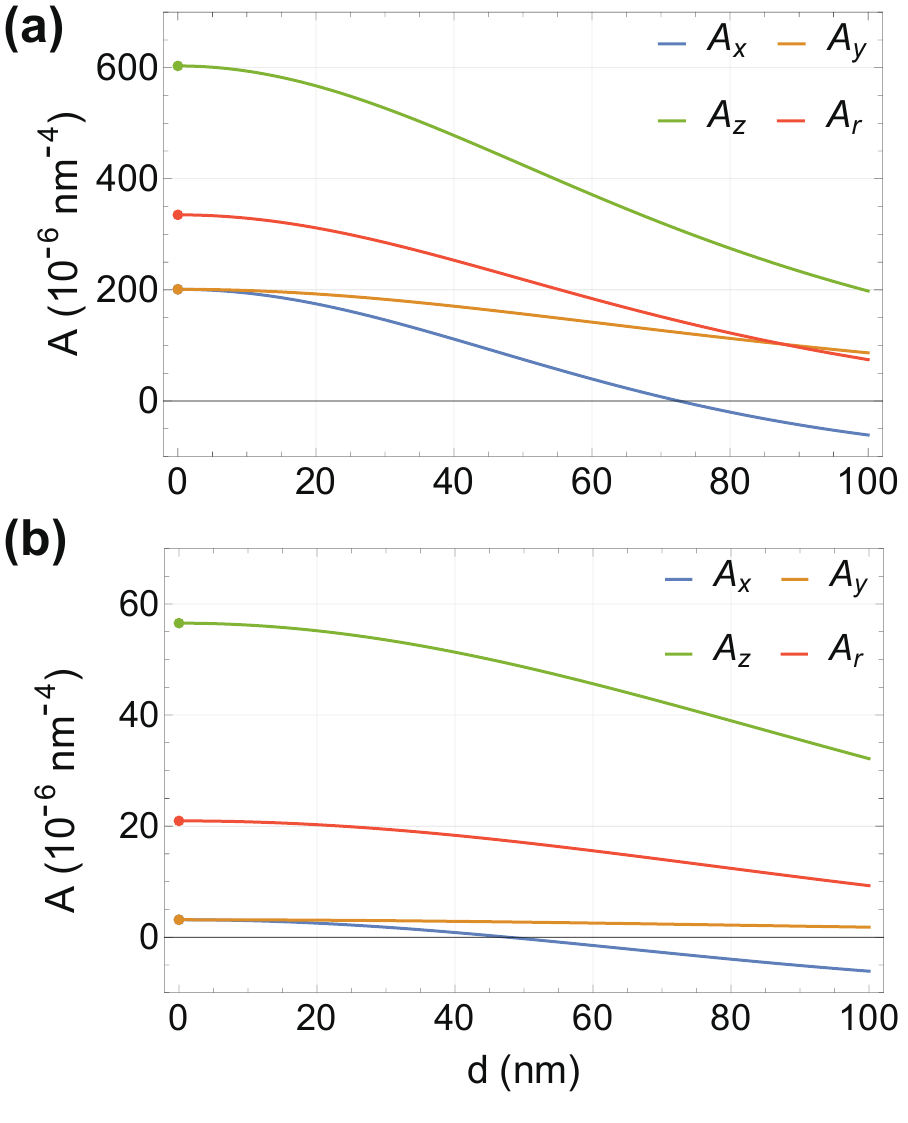}
\caption{Noise correlation strengths $A_x, A_y, A_z, A_r$ between two qubits located at positions $(-d/2,0.0)$ and $(d/2,0,0)$. $A_{x,y,z}$ means that the noise dipoles are in the $x,y,z$ directions, respectively. The noise sources are uniformly distributed on a circular disc of radius $R$ and height $h$ with coordinates $x^2 + y^2 \leq R^2 $ and $z=h$. (a): $R=100$ nm and $h=50$ nm; (b): $R=50$ nm and $h=100$ nm. Note that the $d=0$ limit marked with solid circles is the autocorrelation function for a single qubit, while $d>0$ represents a cross-correlation. Noise dipoles in the $z$-direction are approximately aligned with the qubit-dipole separation vector and they have the strongest effect. $A_r$ (randomly-oriented dipoles) is the average of $A_x,A_y,$ and $A_z$. Of particular interest is the change of sign of $A_x$ at a particular critical separation $d_c$. $d_c \approx 73$ nm in (a) for $R=100$nm while $d_c \approx 47$ nm in (b) for $R=50$nm.}
     \label{fig:A_TLSorient}
\end{figure}

\subsubsection{Results}
We consider APSD (autocorrelations) first.  Then $d\rightarrow 0$ ($\mathbf{R}_1 = \mathbf{R}_2)$ and $D\rightarrow (x^2+y^2+h^2)^{3}$. 
In the case of a circular layer the integrals can be evaluated explicitly.  One finds
\begin{align}
A_x(d=0) &  = 
 \frac{\pi}{4 h^2}
 \frac{R^4}{(h^2+R^2)^2} \\ 
 A_y(d=0) & = 
  \frac{\pi}{4 h^2}
 \frac{R^4}{(h^2+R^2)^2} \\
 A_z(d=0) &  = 
 \frac{\pi}{2 h^2}
 \frac{R^4 + 2h^2 R^2}{(h^2+R^2)^2} \\
 A_r(d=0) & = 
 \frac{\pi}{3 h^2}
 \frac{R^4 + h^2 R^2 }{(h^2+R^2)^2}.
\label{eq:auto4}
\end{align}

We note that $A_z/A_x = 2(1+2h^2/R^2) \geq 2$, since dipole fields are stronger (at a given distance) along the direction of the dipole.  As $R \rightarrow \infty$ $A_z/A_x \rightarrow 2$, the minimum value of the ratio. The qualitative lesson is that, other things being equal, noise from vertically-oriented dipole is stronger than that from horizontally-oriented dipoles. In Sec.~ \ref{subsubsec:orientation} we will show how to make use of this fact.

For the CPSD (cross-correlations) $d \neq 0$. We would usually be interested in dots that are rather close to each other: $d<L$ and $d<h$. The main difference from the previous results is that the numerator for $A_x$ is reduced.  We get $|A_z| > |A_r| >|A_y| > |A_x|$ for $L>h$ and $|A_y| > |A_x| >|A_r| > |A_z|$ for $L<h$.
These inequalities explain the broad trends of the ordering of the $A$'s for the parameter ranges shown in Fig.~\ref{fig:A_TLSorient}. The figure also shows that the effects of changing the orientation depend substantially on the separation between the two quantum dots at which the noise is measured.

\subsubsection{Qualitative Considerations}
\label{subsubsec:qualitative}
The most interesting part of the cross-correlation function is the difference in phase $\gamma$ between the two qubits.  We already noted that $A_x$, necessarily positive for $d=0$, can change sign as $d$ increases. $\gamma$ also changes sign at this point.  

In this subsection we investigate the implications of the sign change for the geometry of the TLS.

\begin{figure}[ht]
\centering
\includegraphics[width=0.48\textwidth]{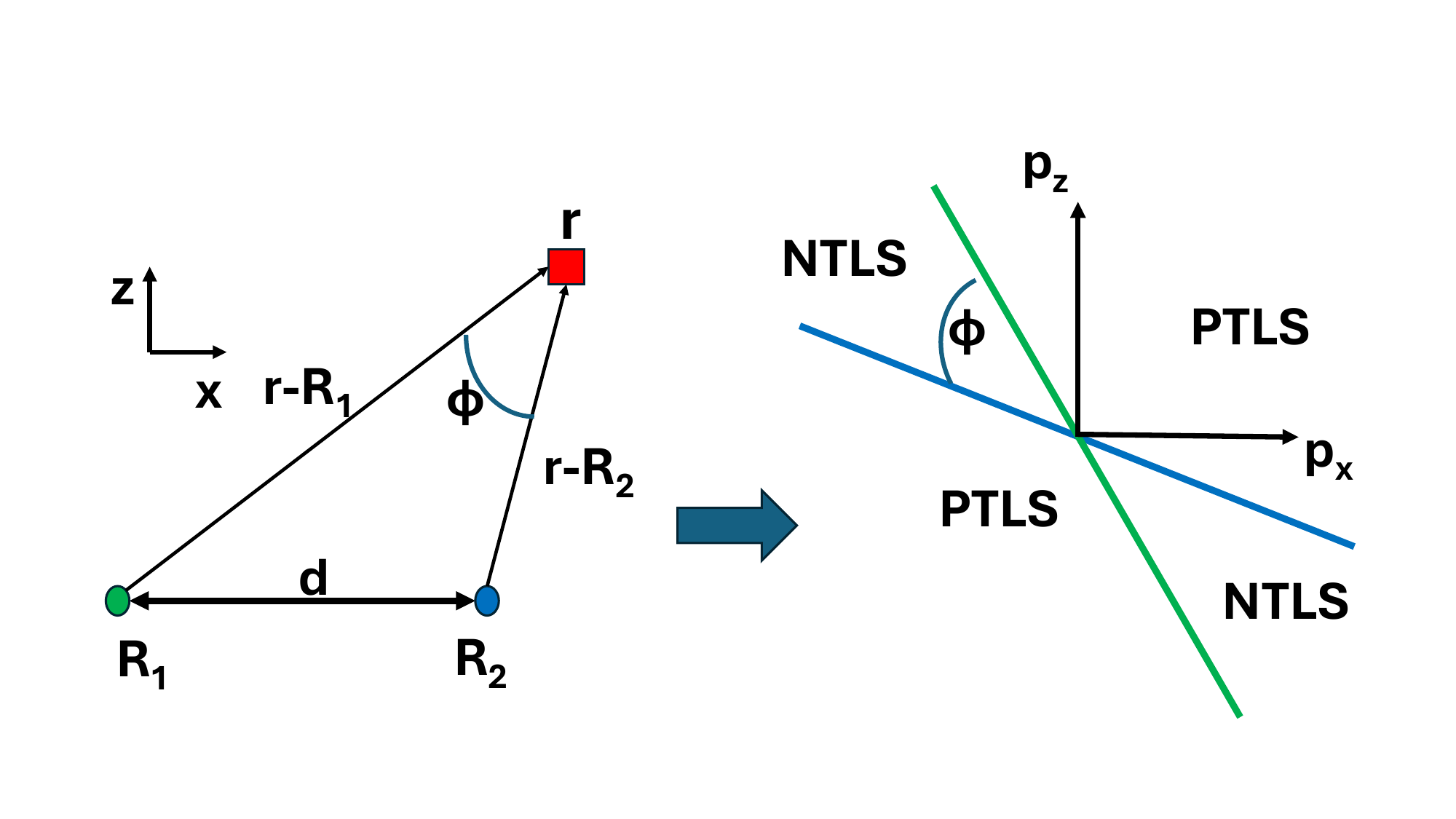}
\caption{Geometry of anticorrelation of noise. On the left are qubit 1 in green at $\mathbf{R}_1$ and qubit 2 in blue at $\mathbf{R}_2$, with a noise dipole in red at $\mathbf{r}$. In the plane determined by the three vectors we can also define the opening angle $\phi$ as shown. The direction $\mathbf{p}$ of the dipole determines whether it produces positive correlation (PTLS) or negative correlation (NTLS). On the right we show the PTLS and NTLS sectors of $\mathbf{p}$-space that follow from Eq.~\ref{eq:neg}. The green and blue planes (shown edge-on) are normal to $\mathbf{r-R_1}$ and $\mathbf{r-R_2}$, respectively. The probability of a TLS in a random position to be NTLS is proportional to $\phi$.  Hence the chance for a TLS to produce negative correlation decreases as it gets further away from the qubits.}
     \label{fig:neg}
\end{figure}

In our model of independent TLS $\gamma$ can only be zero or $\pi$. Consideration of Eq.~\ref{eq:amn} shows that the only way to get a negative sign is through the factor
\begin{equation}
\label{eq:neg}
\mathbf{p} \cdot (\mathbf{r}-\mathbf{R}_1)
\,
\mathbf{p} \cdot (\mathbf{r}-\mathbf{R}_2).
\end{equation}
Converting this vector expression into geometry shows that it 
is negative only if $\mathbf{p}$ has an $x$-component and the dipole is located in the layer $x<|d/2|$ between the qubits.  This is also evident from the numerators in Eq.~\ref{eq:g1}.
These geometrical considerations are illustrated in Fig.~\ref{fig:neg}.

In most of this paper we will focus on a common picture of the TLS as being in a layer that is separated from the qubits: $\rho(x,y,z,\hat{p},\tau)= 0$ when $(x,y,z) = \mathbf{R}_1$ or $(x,y,z) = \mathbf{R}_2$. But the continuum limit also gives valuable information about the APSD when there are TLS close to the qubits: $\rho(x,y,z,\hat{p},\tau) = \rho_0$ when $(x,y,z) = \mathbf{R}_1 = \mathbf{R}_2$.  Then the short-range behavior of $A$ is crucial.  Let us examine the voltage autocorrelation function for a qubit at the origin when the spatial distribution $\rho(x,y,z,\hat{p},\tau)$ for fixed $\tau$ does not depend on the position $(x,y,z)$.  Then we find by inspection of Eqs.~\ref{eq:amn} 
and \ref{eq:vv}
that for all TLS orientations
\begin{equation}
    A(\mathbf{R}_1
    =\mathbf{R}_2=0) \propto 
\rho_0 \int_{r_c}^{\infty} r^{-2} dr,   
\end{equation}
in which a lower cutoff $r_c$ has been inserted because otherwise the integral is divergent.  This indicates that the relatively very few TLSs near $r_c$ dominate the voltage noise.  For electric field noise the situation is even more extreme, since the denominator in the integrand is $r^{-4}$. 

This tells us that in any given frequency range, a small number of TLSs close to the qubit will contribute the majority of the noise, as is sometimes observed.  This also makes clear that a priority for devices is that the near neighborhood of the qubits should be maximally clean. 

\subsection{Cross-correlations}
\label{sec:geometric}
\subsubsection{Introduction}
In the previous section we investigated the dependence of the CPSD of a single TLS on $d$, the separation between the qubits, focusing particularly on the question of the sign of the CPSD.  In this section we analyze what happens when there are multiple TLS with different frequencies.   Again we use geometric analysis but now supplement this with direct simulation of situations where there are a relatively small number of TLS. 

The goal is to develop some intuition about the CPSD in more realistic situations with multiple  TLS.

The CPSD does in fact contain information about the positions and particularly the orientations of the TLS, and even about the position-orientation correlations.  However, this information is obscured by the complicated forms of Eq.~\ref{eq:apsd} and especially Eq.~\ref{eq:ffcorr}.  

The relative phase $\gamma$ of the two qubits is of particular interest.
As shown in Sec.~\ref{subsec:ncf}, $\gamma$ takes only two of the possible values at each frequency: $\gamma (f) = 0$ or $\gamma(f)=\pi$.  This sign is determined by the sign of $S_{ij} (f)$ which involves a summation over the TLS. Let us separate the TLS into two sets: positively correlated TLS (PTLS) or negatively correlated TLS (NTLS).  A PTLS (NTLS) at position $\mathbf{r}$ with dipole moment has a positive (negative) value for the product $(\mathbf{p}\cdot\hat{\mathbf{u}}_{1})(\mathbf{p}\cdot\hat{\mathbf{u}}_{2})$,
where $\hat{\mathbf{u}}_{i} = (\mathbf{r}-\mathbf{R}_i)/|\mathbf{r}-\mathbf{R}_i|$. This classification can be interpreted geometrically in two-qubit QD devices. We consider two planes that pass through the origin whose normal vectors are parallel to  $\mathbf{u}_1$ and  $\mathbf{u}_2$, respectively.  Each plane divides $\mathbf{p}$-space into two half spaces, one with $\mathbf{p}\cdot \mathbf{u}_i<0$ and one with $\mathbf{p}\cdot \mathbf{u}_i>0$. The TLS is a PTLS if the product of the signs is positive and a NTLS if the product is negative.  This argument is shown visually in Fig.~\ref{fig:neg}. We deduce from the figure that the chance of a TLS being NTLS decreases as the distance from the qubits increases. Also, $\mathbf{p}$ must have a component along the separation vector of the qubits, here the $x$ direction. 

\begin{figure}[ht]
\centering
\includegraphics[width=0.48\textwidth]{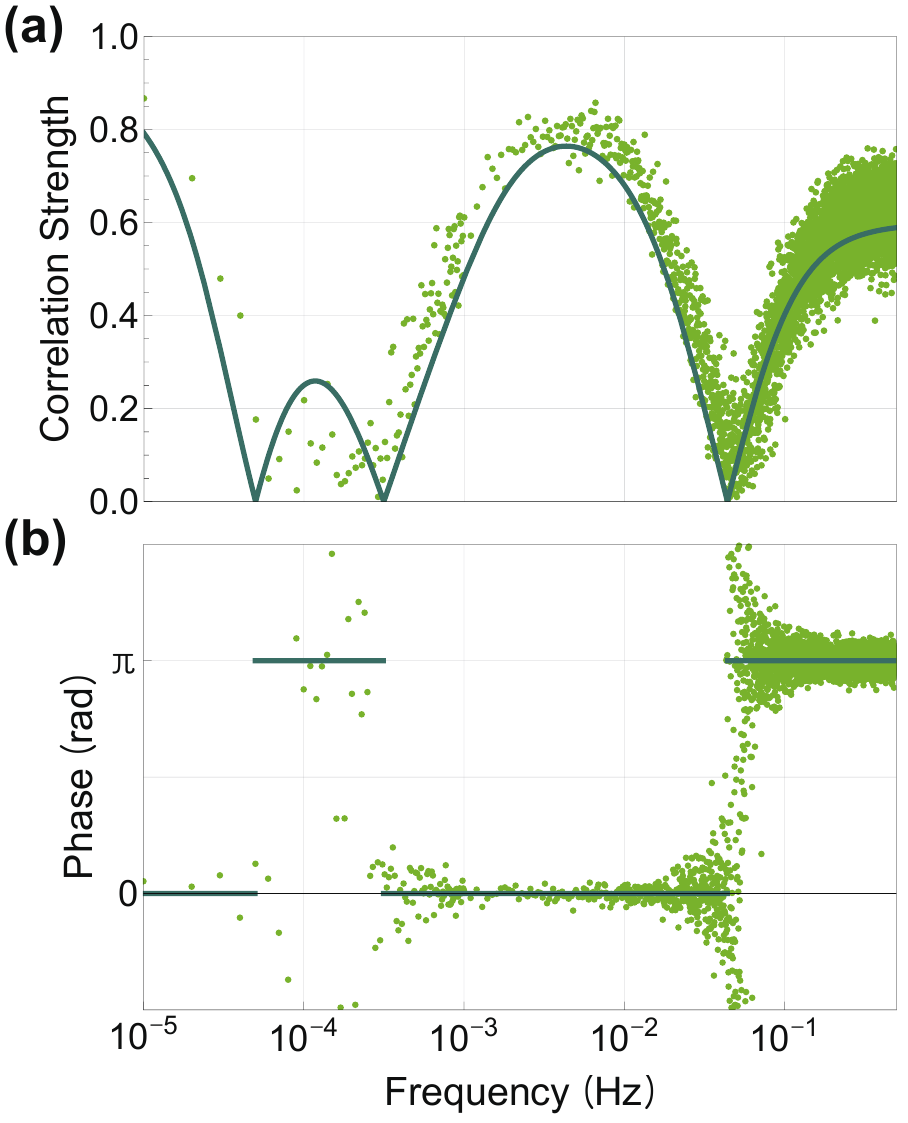}
\caption{An example of CPSD with 10 noise dipoles distributed on a 2D plane.  (a) shows the magnitude of the CPSD and (b) shows the phase $\gamma$. Lines show results of the analytic expression for CPSD, while dots are the results from a Fast Fourier Transform (FFT) of the time series generated by a Poisson process for the TLS.  For clarity, the number of points shown in the figure for frequencies higher than $10^{-3}$ Hz is reduced. Changes in the phase $\gamma$ from 0 to $\pi$ correspond to zeros in the correlation. This example shows that noise leakage in the FFT produces deviations from the exact result, and give phases with $\gamma \neq 0, \pi$.}
     \label{fig:CPSD_ex}
\end{figure}

\subsubsection{Small Systems}
\label{subsubsec:computation}

We now turn to the consideration of small ($N\leq 10$) model systems of TLS to understand the CPSD. We compare two different methods of calculating the CPSD  $C_{12} (f)$ in Fig.~\ref{fig:CPSD_ex}. As an example, we generate a single configuration of 10 TLS drawn from a spatially  uniform distribution on a 2D plane which might be an interface between an oxide layer and a gate layer ($z = 72$ nm).  The two qubits are located in a 2DEG with Cartesian coordinates, QD1 = $(-50, 0, 0)$ nm and QD2 = $(50, 0, 0)$ nm, respectively.  The dipole orientations are in the $x$-$y$ plane with the angle from the $x$-axis uniformly distributed. The switching rates of the TLSs are sampled from a log-uniform distribution over ($10^{-5}$, 1) Hz to ensure their APSDs have a $1/f$-like form over this frequency range.  
The green data points are averages of the Fourier transform of 100 simulated time series for the duration of $10^5$ s with sampling frequency of 1 Hz.

The first computation method uses the exact analytic formula in Eq.~\ref{eq:CPSD} to get the solid lines. The second is from 100 time series that assume a Poisson process for each TLS and a fast Fourier transform (FFT) is then performed to generate the discrete points.  The results from the two methods for the magnitude and phase of the FFT as a function of frequency are shown in Fig.~\ref{fig:CPSD_ex}.

The magnitude has a series of peaks associated with 
one or a few TLS of the appropriate frequency. There is anticorrelation in some frequency ranges, and since in the exact solution the CPSD is real, this means a change of sign and zeros at certain discrete frequencies.  As shown above, the statistical independence of the TLS in the model under discussion implies that $\gamma = 0,\pi$ in an exact calculation, and indeed the solid lines are in agreement with this. 

The results from the FFT show characteristic deviations from the exact results. This phenomenon is known generically as spectral leakage  \cite{cerna2000fundamentals}. 
The deviations between the two computations are due to two important sources.  

1. The signal is sampled for a finite time.  This produces the well-known and relatively trivial broadening of structures in the FFT, evident particularly in the larger peaks in the magnitude but present also in the phase plot.  
 More interesting from our point of view is that the averages $\langle s_m(t)  s_n(0) \rangle$ in Eq.~\ref{eq:apsd} for $m \neq n $ only vanish when the average is taken over a long enough time.  This produces some spurious correlations among the TLS. 
 
 2. The time series has a finite sampling time interval $\tau$, which will produce errors at frequencies greater than Nyquist frequency, $1/2\tau$ = 0.5 Hz in these simulations.  

The phase results for the the analytic expression and the FFT for the Poisson process agree fairly well except near the frequencies when there is a transition between the two phase values 0 and $\pi$.   This is due to the fact that when the correlation strength is close to zero, small changes in the complex number that represents $C_{12}(f) $ can drastically change $\gamma$.  
These results show that the phase can be different for 0 or $\pi$ for rather trivial reasons. But there is an additional important physical point, namely that when deviations of $\gamma$ from 0 and $\pi$ are actually observed experimentally, this may be due to the TLS not being completely independent, that is $\langle s_m(t)  s_n(0) \rangle \neq 0$ for $m \neq n$. This would indicate TLS-TLS interactions, which are not present in our simulations. Our results show that although such interactions may be present in experiments, one must take care to disentangle such genuine physical effects from effects coming from finite sampling windows.

\subsubsection{Orientation Effects}
\label{subsubsec:orientation}
Though the data for a single TLS configuration in Fig.~\ref{fig:CPSD_ex} are qualitatively representative of the frequency dependence of CPSD in systems with just a small number of TLS, the details are quite sensitive to the specific realization of the TLS.  In particular, the orientations of the TLS make a considerable difference, as we would expect from Eq.~\ref{eq:gdef}. This implies that a statistical analysis that focuses on the orientation can shed light on the physical nature of the TLS.

To analyze the data in a systematic way, we simulated $10^4$ samples with the analytic formula in Eq.~\ref{eq:voltage} for five orientation models with the TLS being distributed on a 2D plane. 

 The dipole vector is represented as $\mathbf{p} = p(\sin\theta \cos\phi,\sin\theta \sin\phi,\cos\theta)$. The orientation models are specified by the distributions of polar angle ($\theta$) and azimuthal angle ($\phi$).  The five orientation models that we investigated are: Fully R, Hor.(R), Hor.(x), Hor.(y), and Ver.(z) models.  The fully R model is the fully random orientation model where all TLS have random orientations on a 3D unit sphere ($\theta \in [0,\pi)$ and $\phi \in [0,2\pi)$): $\theta = \arccos{u}$, with $u$ sampled uniformly from $[-1, 1]$. The Hor.(R) model has TLS randomly oriented in the $x-y$ plane ($\theta=\pi/2$ and $\phi$ uniformly distributed in $[0,2\pi)$). The Hor.(x) and Hor.(y) models have the TLSs all pointing along the $x$ direction: ($(\theta,\phi)=(\pi/2,0)$) or all along the $y$ direction: $(\theta,\phi)=(\pi/2,\pi/2)$, respectively.  Ver.(z) model has the TLSs oriented vertically along the $z$ direction: $(\theta,\phi)=(0,0)$.
 
In a 2D model, the TLSs are assumed to be at the (very thin) interface between the oxide layer capping semiconductor surface and a metallic gate. In a 3D model they would in an oxide layer of finite thickness. We found that the 3D case was qualitatively similar to 2D, and thus do not present 3D results.  

\begin{figure}[ht]
\centering
\includegraphics[width=0.48\textwidth]{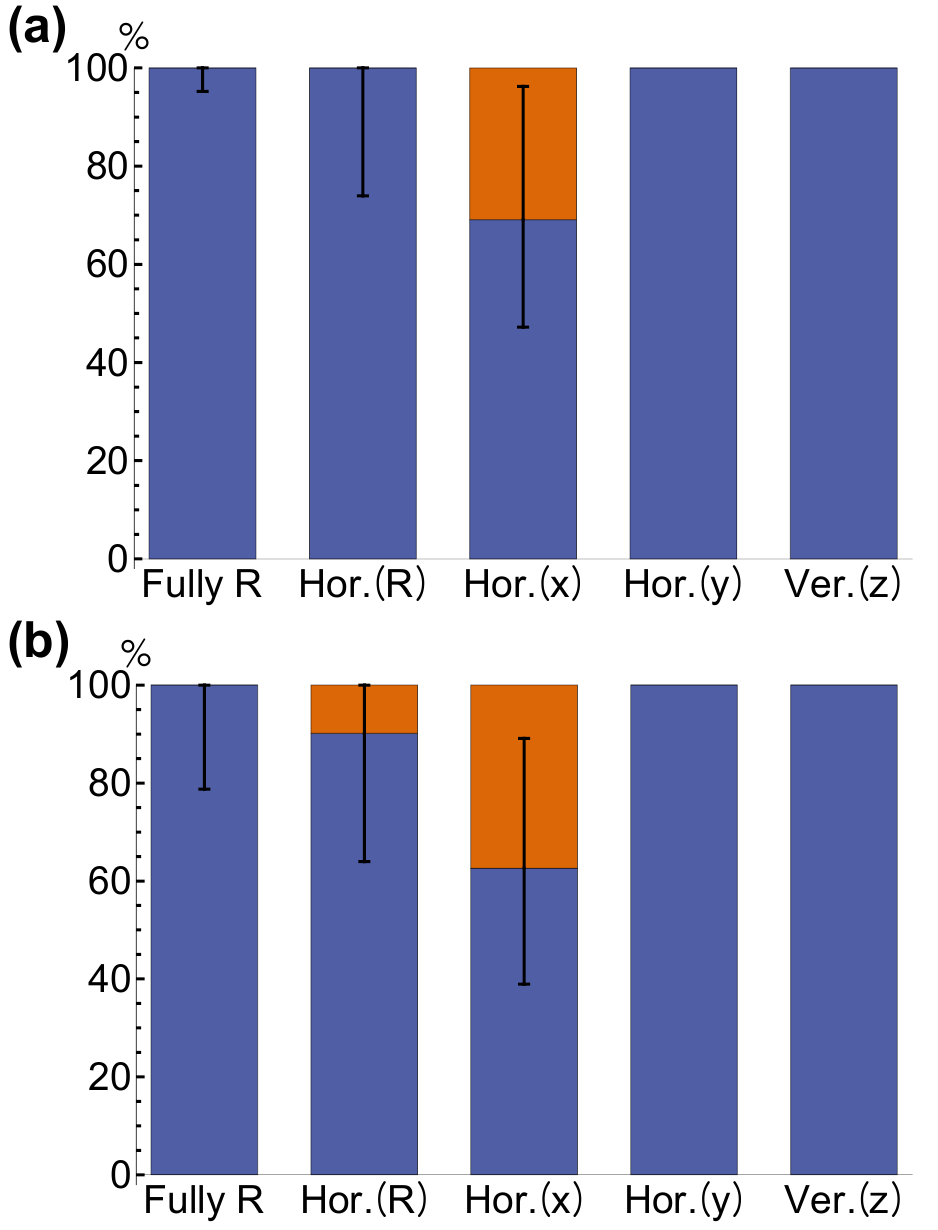}
\caption{Weighted percentage of $0$ and $\pi$ phase of the CPSD for different orientation models of 10 noise dipoles distributed on a 2D layer. The percentages are shown in (a) for the layer at a height of 72 nm and in (b) a height of 50 nm.  Blue (Orange) bar denotes the percentage of $0$ ($\pi$) phase.  Medians are marked as the border lines between the bars and interquartile ranges as the error bars.  No error bars mean that there is no uncertainty over the simulated samples.  Labels for dipole orientations are given as Fully R: fully random, Hor.(R): random in an $xy$ plane, Hor.(x): TLSs aligned in the $x$ direction, Hor.(y): TLSs aligned in the $y$ direction, Ver.(z): TLSs aligned in the $z$ direction.  The densities of noise dipoles used in the simulations are $1.11 \times 10^{10}$ cm$^{-2}$. The fact that Hor.(x) gives significant anticorrelation is due to the TLS dipoles having strong $x$ components. Hor.(R) and Fully R have progressively weaker $x$ components and Hor.(y) and Ver.(z) have none. The anticorrelation is stronger for closer noise sources.}
     \label{fig:10ND_phs_main}
\end{figure}

\begin{figure}[ht]
\centering
\includegraphics[width=0.48\textwidth]{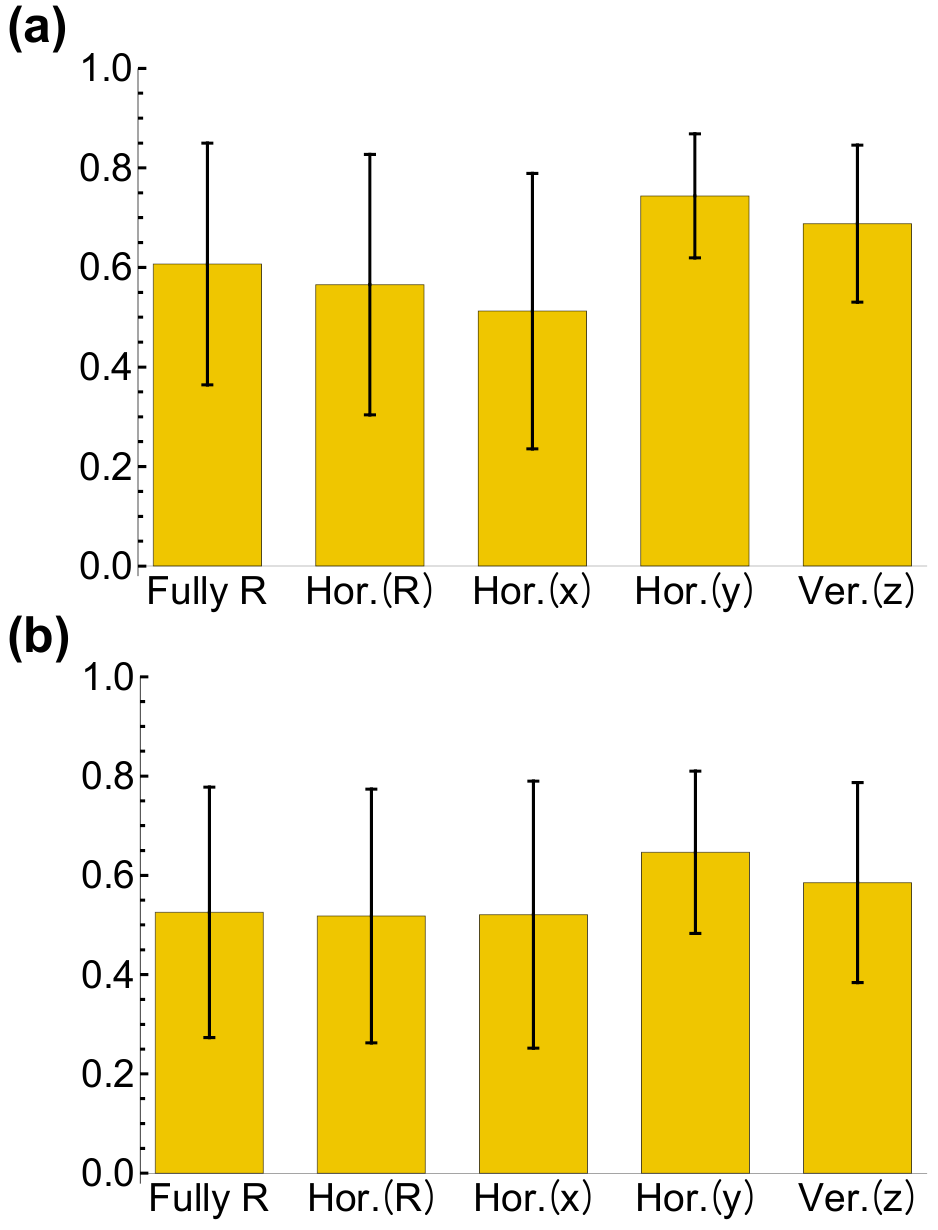}
\caption{Correlation strength for different orientation models of 10 noise dipoles distributed  on a 2D layer. The correlation strengths between the potential of QD1 and QD2 are shown in (a) for the 2D layer at a height of 72 nm and in (b) for a height of 50 nm.  Yellow bars (error bars) denote the mean (standard deviation) of correlation strength.  The means and standard deviations are obtained by averaging over 4 frequencies at $10^{-4}$, $10^{-3}$, $10^{-2}$, and $10^{-1}$ Hz and over samples.  Labels for TLS orientations are given as Fully R: fully random, Hor.(R): random in an $xy$ plane, Hor.(x): TLSs aligned in the $x$ direction, Hor.(y): TLSs aligned in the $y$ direction, Ver.(z): TLSs aligned in $z$ direction.  The densities of noise dipoles used in the simulations is $1.11 \times 10^{10}$ cm$^{-2}$. Anti-correlation in the Hor.(x) and (more weakly) in the Hor.(R) and Fully R models give weaker average correlation.}
     \label{fig:10ND_corr_main}
\end{figure}

The positions of quantum dots for Figs. \ref{fig:10ND_phs_main} and \ref{fig:10ND_corr_main} are $(x,y,z) = (-50, 0, 0)$ and $(50, 0, 0)$ nm, respectively.  The noise dipoles are randomly sampled from a uniform distribution over $-150 < x,y < 150$ nm with $z = 72$ nm for the 2D distribution of dipole positions.  The $z$ coordinate for the dipoles for the 2D case is the position where depletion gates are located in the device.  These inter-dot spacing and $z$ coordinates are taken from Ref.~\cite{Takeda:2016p1600694}, which used the same device in Refs.~\cite{yoneda2020quantum, yoneda2023noise}.  
The dielectric constant $\epsilon/\epsilon_0$ used in the simulation is 11 where $\epsilon_0$ is the vacuum permittivity.  The dipole magnitude is $p_0 = 48$ Debye $\approx 1 |e|$-nm, which is a representative value with elementary charge $e$ hopping between sites separated by 1 nm~\cite{choi2024interacting}. Note, however, that $p_0$ and $\epsilon$ cancel out in the normalized CPSD.  

The results for the phase of and the strength of the CPSD are shown in Figs. \ref{fig:10ND_phs_main} and \ref{fig:10ND_corr_main}.
The switching rates are again sampled from a log-uniform distribution over $f \in (10^{-5}, 1)$ Hz. 

Here, we focus on statistical tendencies that show up when we look at the entire frequency range.  To this end, we introduce a weighted percentage to account for the ratio of $0$ and $\pi$ phases of CPSD for the frequency range given above. The weight factor is $1/f$, which allocates equal percentage for each decade of frequency.  This weight factor also conforms well with the visualization of experimental data whose FFT results are typically shown in a log-scale frequency axis.  For instance, the weighted percentages for 0 and $\pi$ phase in Fig. \ref{fig:CPSD_ex} are $56.9\%$ and $43.1\%$. In Fig. \ref{fig:10ND_phs_main}, we choose to use median values of the percentages of 0 and $\pi$ phase denoted by the borders of blue and orange solid bars respectively. The error bars are located at interquartile boundaries, a choice that was made to emphasize the fact that the distributions are usually skewed. These border lines are not shown in the cases where the percentage of phase $0$ or $\pi$ is close to 100\%.  

Panels (a) and (b) of Fig.~\ref{fig:10ND_phs_main} exhibit the phase of the CPSD for the electric potential. They  compare TLS distributed on a 2D layer at heights $z = 72$ nm and $z = 50$ nm, respectively.  

Our arguments in Eq.~\ref{eq:neg} and the ensuing discussion show that the phase for the CPSD in this layer model is always zero for the Hor.(y) and Ver.(z) models, since the dipoles have no $x$ component. The Hor.(x) then exhibits the strongest anticorrelation, and it becomes progressively weaker as we pass to the Hor.(R) and Fully R models.

The correlation strength is closely related to the phase as illustrated in Fig. \ref{fig:10ND_corr_main}.  Overall, we found that as in the continuum limit the correlation strength for electric potential increases from Hor.(x), Fully R, Ver.(z), to Hor.(y) model with the simulation parameters, which agrees with the results in Fig. \ref{fig:10ND_corr_main}(a) and (b) even though only a quite small number of TLS is used in the simulation.  It is striking that the main difference in the correlation strengths can be traced back entirely to the anticorrelation.  Where anticorrelation is present, the average correlation is less due to cancellations. 

There are three qualitative lessons that emerge from these data.

1. The orientation is very important.  Dipoles oriented in the direction that connects the two dots is necessary to get the phase $\gamma = \pi$. The only orange in the figures (strongly negative correlation) is for the Hor. (x) model.  Some negative correlation is present for the Hor. (R) model, in which an $x$-component is present but not dominant in the dipoles.  

2. When the distance from the TLS to the qubits is greater than the inter-qubit distance then $\gamma = 0$ is strongly favored.  This is why blue predominates in Fig.~\ref{fig:10ND_phs_main}(a). Experimental observation of strongly positive  corelation is indicative of a layer model, or at least a model in which most of the noise is coming from TLS whose distance to the qubits is greater than the inter-qubit separation. 

3. The anisotropy of the anticorrelation gives a rather direct way to measure the in-plane orientation of the TLS. One can do an experiment with four dots at the vertices of a square, and measure all six pair correlation functions.  Anticorrelations that appear give evidence of TLS orientations along the direction of the ``bond''.  

\section{Probabilistic Analysis}
\label{sec:bayes}

\subsection{Bayesian formalism}
\label{subsec:bayes}

The conclusion of Sec.~\ref{sec:under} was that the goal of determining all of the parameters of the TLS is too ambitious.  We must moderate our expectations.  Our aim in this section will be to create a rational assignment of likelihood to regions of the configuration parameter space.
This goal suggests a Bayesian analysis of the configuration space \cite{garnett}. A good introduction to this type of anaylsis in a similar context is given by Gut\'ierrez-Rubio \textit{et al}., who use it to determine correlation functions \cite{gutierrez}.  

The terminology for this section is as follows.  A configuration is a complete specification of the number of TLS dipoles and their positions, magnitudes, relaxation times and orientations. A model $\mathcal{M}_j$ is a subset of all the configurations.  An example of a model $\mathcal{M}_j$ would be a subset in which all TLS have the same fixed magnitude $p$. Each subset represents a hypothesis about the device. We label these hypotheses by $j$. $j$ can be continuous if, for example, we consider a range of values of $p$.

Our goal is to evaluate hypotheses given some experimental data.   Because of the size of the parameter space it is  necessary to narrow down the range of hypotheses about the TLS. In this work
we focus on models $\mathcal{M}_j$ where $j$ represents a choice of the total number $N_T$, $p = |e|\ell$, the magnitude of all dipoles, and the orientation. Here $|e|$ is the electron charge so $\ell$ is a measure of the length of the dipole, while the orientation $i_{or}$ will be limited to vertical [$i_{or}=$Ver.(z)], horizontal [$i_{or}=$Hor.(R)], and fully random [$i_{or}=$Fully R]. Because of this focus we also write $P(\mathcal{M}_j) = P(\mathcal{M}(N_T, \ell, i_{or}))$ when we wish to bring attention to the particular physical features of some of our results. 

Our first step is to assign a probability $P(\mathcal{M}_j)$ to each $\mathcal{M}_j$.
To each configuration we assign a ``prior'' probability $P_0(\mathcal{M}_j)$. In this paper we also assume that the TLS parameters are statistically independent so $P_0(\mathcal{M}_j)$ is a product.  For the positions, we suppose that the TLS are confined to a layer, perhaps an oxide layer or metal-semiconductor interface.  Within this layer $P_0$ is a uniform distribution in the position variables. For the Hor.(R) model the azimuthal angle of the dipole is uniformly distributed in the interval $[0,2 \pi]$. For the Fully R model the cosine of the polar angle is also uniformly distributed in the interval $[-1,1]$.

The data $\mathbf{x}$ is a set of numbers that are the results of an experimental measurement (in our case the APSD or CPSD). There may also be experimental error.  In this case we define $p(\mathbf{x})$ to be the probability density of obtaining that measurement given an error model.   

The measurement set $\mathbf{x}$ is then used to update the probability $P_0$ according to Bayes' rule:
\begin{equation}
\label{eq:bayes}
P\left(\left. \mathcal{M}_j \right| \mathbf{x} \right) = 
 \frac{p\left( \left. \mathbf{x} \right| \mathcal{M}_j\right) P_0\left( \mathcal{M}_j \right)}{p\left( \mathbf{x} \right)}.
\end{equation}
Here $P\left( \left. \mathcal{M}_j \right| \mathbf{x} \right)$ is the probability of $\mathcal{M}_j$ given $\mathbf{x}$.

The probability of obtaining a particular set of measurements $p\left( \mathbf{x} \right)$ is 
\begin{equation}
P\left( \mathbf{x} \right) =
	\sum_j p\left( \left. \mathbf{x} \right| \mathcal{M}_j\right) P\left( \mathcal{M} \right).
\end{equation}
The validity of this expression depends on how complete the set of configurations $\mathcal{M}_j$ is; if the set is incomplete then there will be an out-of-model error, leading to an overestimate for the probabilities $P\left(\mathcal{M}_j\right)$.

If our measurements do have errors, then we define a set $X$ containing all points $x'$ that fall within the error bars for each measurement.  The appropriate generalization of Eq.~\ref{eq:bayes} is 
\begin{align}
P\left(\left. \mathcal{M}_j \right| X \right) &= \int_{\mathbf{x}' \in X} \dd{\mathbf{x}'} p\left(\left. \mathcal{M}_j \right| \mathbf{x}' \right) p(\mathbf{x}').
\end{align}

The likelihood functions $p\left( \left. X \right| \mathcal{M}_j\right)$ are in general nontrivial to calculate analytically, but can be estimated via a Monte Carlo method.
For each configuration $\mathcal{M}$, we generate a large number $N$ of measurements $\mathbf{y}$, and count the number of measurements $n_{\mathrm{success}}$ that fall within $X$ using Eqs. \ref{eq:voltage}-\ref{eq:CPSD}. 
Then
\begin{equation}
	P\left( \left. X \right| \mathcal{M}_j\right) \approx \frac{n_{\mathrm{success}}}{N},
\end{equation}
and the approximation improves as $N$ gets larger.

The comparison of the posterior probabilities $P\left( \left. \mathcal{M}_j \right| X \right)$ for different $j$ then allows us to assign probabilities for the varying dipole counts, magnitudes and orientations.
If an experiment is repeated, it is not expected that the positions of each TLS will be the same in a new device, but as long as the set $\mathcal{M}_j$ for all $j$ stays the same, the Bayesian process can be iterated, using the previous experiment's probabilities $P\left( \left. \mathcal{M}_j \right| X \right)$ as the priors for another analysis.

\begin{table*}
	\begin{tabular}{c | c c c | c}
		Dipole Magnitude: & 10 & 100 & 1000 & Total \\
		\hline
		1 dipole & 0.762 & \textbf{0.218} & 0.019 & 0.9999 \\
		10 dipoles & $4.023 \times 10^{-7}$ & $4.023 \times 10^{-7}$ & $4.023 \times 10^{-7}$ & $1.207 \times 10^{-6}$
	\end{tabular}
	\caption{
		The probabilities predicted for six possible configuration parameters from a single dot.
		The actual noise source was a single TLS with magnitude $100$, corresponding to the cell marked in bold.
		For this case, the dipole count was correctly predicted to be $1$, rather than $10$.
		However, the magnitude was incorrectly predicted, with a 76\% probability assigned to a dipole magnitude of $10$.
		A perfect predictor would assign a probability of $100\%$ to the correct answer, and $0$ to everything else.
        The high Brier score for this case of 1.19 indicates low accuracy or the results obtained.
	}
	\label{tab:exampleDot1Probabilities}
\end{table*}

\begin{table*}
	\begin{tabular}{c | c c c | c}
		Dipole Magnitude: & 10 & 100 & 1000 & Total \\
		\hline
		1 dipole & 0.302 & \textbf{0.545} & 0.153 & 0.9999 \\
		10 dipoles & $8.278 \times 10^{-6}$ & $8.278 \times 10^{-6}$ & $8.278 \times 10^{-6}$ & $2.48 \times 10^{-5}$
	\end{tabular}
	\caption{
		The probabilities predicted for six possible configuration parameters from two dots, for the same mock configuration analyzed in \cref{tab:exampleDot1Probabilities}.
		The actual noise source was a single TLS with magnitude $100$, corresponding to the cell marked in bold.
		The probability has increased for the correct answer, mostly because of the lower probability assigned to the incorrect magnitude $10$.
		The Brier score for this case has decreased to $0.321$, showing that the second dot has indeed increased our accuracy.
	}
	\label{tab:exampleDot2Probabilities}
\end{table*}

\begin{table*}
	\begin{tabular}{c | c c c | c}
		Dipole Magnitude: & 10 & 100 & 1000 & Total \\
		\hline
		1 dipole & 0.208 & \textbf{0.776} & 0.016 & 0.9999 \\
		10 dipoles & $8.18 \times 10^{-7}$ & $4.62 \times 10^{-10}$ & $4.20 \times 10^{-12}$ & $8.18 \times 10^{-7}$
	\end{tabular}
	\caption{
		The probabilities predicted for six possible configuration parameters from a single dot.
		In addition to the sample analyzed in \cref{tab:exampleDot1Probabilities}, we generate a second sample.
		Now, the results for the correct answer are noticeably better, and the Brier score has dropped to $0.094$.
		This suggests that measuring on two separate samples with newly generated noise dipoles provides far more information than adding a second dot to a single measurement.}
        \label{tab:three}
\end{table*}

\subsection{Brier score}
\label{subsec:brier}
The end result of the Bayesian method as described above produces the probability $P\left( \left. \mathcal{M}_j \right| X \right)$. Better measurements will clearly produce a better $P$ and we need a quantitative confidence measure for $P$ to understand how this happens. This is of crucial importance in the design of experimental tests \cite{lichtenstein1982}.
There is no single universally accepted measure, but the most common is the Brier score ($B$) \cite{brier1950,murphy1972}.  This is defined as

\begin{equation}
\label{eq:BS}
	B = \sum_{j} \left(P\left(\mathcal{M}_j \mid y\right) - \delta_{j} \right)^2,
\end{equation}
where the index $j$ ranges over the configuration parameter options and $\delta_j$ is an indicator that equals $1$ when $j$ is the index of the correct response and zero otherwise.
$P(\mathcal{M}_j \mid y)$ is, as above, the predicted probability given our data.
$B$ ranges from $0$ to $2$, with a lower score indicating greater probability assigned to the correct answer and low probabilities assigned to incorrect answers.

A perfect guess, assigning probability $1$ to the right answer and $0$ to everything else, gives a score of exactly $0$.
A no-confidence guess assigning an equal probability to all options leads to $B \approx 1$.

We repeat the same Bayesian analysis as above, but now for a variety of different simulated noise sources.
We begin by creating a distribution of TLS dipoles whose random telegraph noise generates a random time-dependent electric potential.
The dipoles are chosen to be distributed uniformly within a $200 \times 300 \times 25$ nanometer layer, with bottom face $\SI{50}{\nano\meter}$ above the plane of the dots. $10^5$ samples were generated. After binning, this gives a range of about four decades of frequency to analyze.

We begin by evaluating the electric potential noise from a single known dipole with known orientation.
Our single ``actual'' dipole is created with dipole magnitude $100$ (with arbitrary units).
In the test, three magnitudes are chosen: $10$, $100$ and $1000$.
Additionally, we test dipole counts of $1$ and $10$.
This lets us determine whether a single dipole of magnitude $100$ is confused with ten dipoles with magnitude $10$, for example. This gives 6 configuration parameter options, so the index $j$ in Eq.~\ref{eq:BS} ranges from 1 to 6.

We focus on whether the Bayesian analysis is able to extract information about the aggregate TLS distribution data, here the number and dipole magnitude, and do not try to estimate the exact parameters of each individual TLS for this analysis.

The evaluated probabilities for this trial for a single dot are shown in \cref{tab:exampleDot1Probabilities}.
In this case, the dipole count is correctly predicted to be $1$, but a magnitude of $10$ is given a higher probability than the correct magnitude $100$.

For the data shown in \cref{tab:exampleDot1Probabilities}, the Brier score is $1.19$, a value that is worse than the no-confidence guess.
Much of the penalty in the score comes from the confidence given to the incorrect magnitude of $10$.
If we repeat the same mock experiment, with the dipole located in the same place, but with measurements taken at \emph{two} dots, our results improve, as displayed in \cref{tab:exampleDot2Probabilities}.
Compared to the single dot case, our probability has increased for the correct answer while decreasing for the case of a magnitude of $10$. The confidence has increased: now $B=0.321$

An additional benefit of the Bayesian method is that it provides a  natural way to handle measurements of multiple samples.
The Bayesian analysis of one sample gives probabilities that can serve as priors for future samples.
Because the TLS distribution is effectively random, there is little reason to expect devices to have identical noise profiles, even if manufactured by the same process with the same design.  However, under the assumption that the noise sources are sampled from the same distribution, sampling multiple devices constrains the distribution more effectively than using only one.

To simulate this, we create a second sample and subject it to the same analysis, using Bayes' theorem to combine the samples into a single prediction.
This means we keep the same dipole magnitude, count and orientation, but generate a new configuration with different positions.

To illustrate this, we add a second sample to the one already treated in  \cref{tab:exampleDot1Probabilities}. Again, just one dot is used for measurement.  The results are shown in \cref{tab:three}.
They demonstrate that an additional sample greatly improves our confidence in the conclusions: the Brier score is now $B = 0.094$.

We conclude that the confidence in the probability is dramatically improved when the amount of data increases. This is hardly surprising. It is somewhat unexpected that two samples, even if measured with only one dot, are more effective than one sample with two measurement dots.
But what is most important about our analysis is that the Brier score puts this whole process of measurement design on a quantitative basis.  

\subsection{Analysis of Experimental APSD} \label{subsec:bayesapsd}

\begin{figure}
	\includegraphics[width=250px]{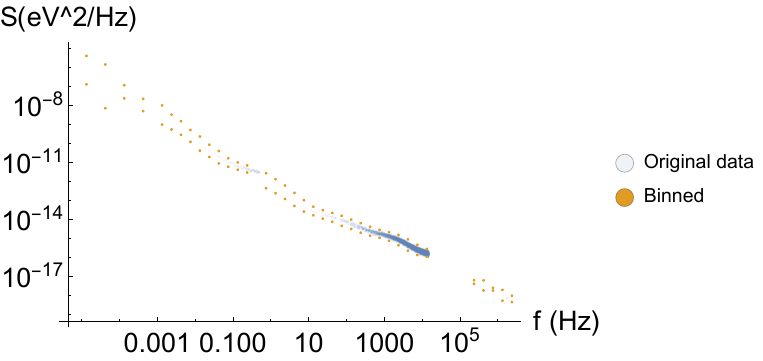}
	\caption{The measured noise from \textcite{connors2022charge}, along with the binned ranges used in the Monte Carlo simulation. A successful Monte Carlo configuration is one where the simulated noise passes between the upper and lower range for each bin.} 	\label{fig:connorsDataPlot}
\end{figure}

We apply this Bayesian analysis to the data in Connors \textit{et al.}~\cite{connors2022charge}.
The data set $X$ is obtained by taking the measured values of $S(f)$ from Fig.~\ref{fig:connorsDataPlot}, binning them, and then including standard deviations for each bin.  This gives a set $X$ of manageable size.

\begin{figure}
	\includegraphics[width=250px]{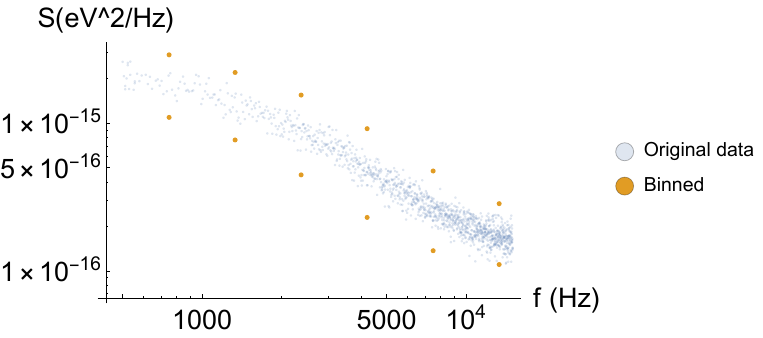}
	\caption{A portion of the measured noise from \textcite{connors2022charge}, along with the binned ranges used in the Monte Carlo simulation. A successful Monte Carlo configuration is one where the simulated noise passes between the upper and lower range for each bin. Only a selection is presented for visual clarity, but the Bayesian results are calculated using the full data. }	\label{fig:connorsplotrestrictedbins}
\end{figure}

The cost function is defined so that the deviation allowed in our Monte Carlo step is where the simulated noise power $S_{i, sim}$ falls within the range $S_i(f) \pm 3 \Delta S_i(f)$ for each frequency $f$.
The allowed range and the bin width are chosen to balance computation speed and accuracy.
Widening the range allows for more Monte Carlo successes, but risks allowing matches that are unrepresentative of the underlying TLS distribution.
A portion of this data, with bins, is shown in Fig.~\ref{fig:connorsplotrestrictedbins}.

The probabilities $P(\mathcal{M}(n_T, \ell, i_\mathrm{or}))$ are calculated using the method described above.
The values of the parameter $N_T$ are taken as a geometric series with 13 members starting at 1 and ending at 177. The values of $p$, the dipole moment magnitude ranges are taken to range uniformly over 2 orders of magnitude.  Treating such a wide range of possible parameters would not be feasible with standard optimization methods.

The TLS count and dipole moment magnitude are represented in terms of a dipole areal number density $n_T$ and a length $\ell$.
The results $P(\mathcal{M}(n_T, \ell, i_\mathrm{or}))$ are summed over orientation, with $P(\mathcal{M}(n_T, \ell))$ representing the sum of the probabilities over all $i_\mathrm{or}$.
\begin{equation}
	P(\mathcal{M}(n_T, \ell)) \equiv \sum_{i_\mathrm{or}} P(\mathcal{M}(n_T, \ell, i_\mathrm{or}))
    \label{eq:probdensityplotdefinition}
\end{equation} 

\begin{figure}
	\includegraphics[width=250px]{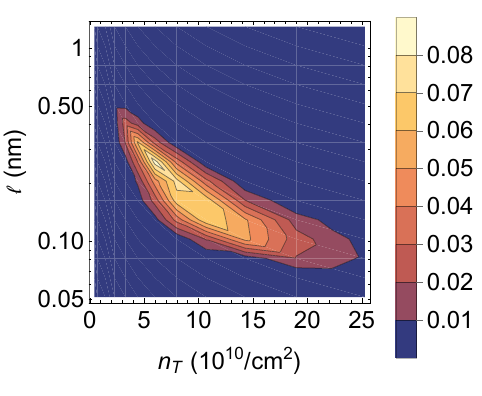}
	\caption{
    Color map plot of the probability $P(\mathcal{M}(n_T, \ell))$ for different TLS density $n_T$ and dipole length $\ell$, as obtained from our Bayesian analysis given the measured PSD from \textcite{connors2022charge}.
    Results are summed over the three orientations tested, as described in Eq.~\ref{eq:probdensityplotdefinition}.
    The overall shape of the distribution represents the tradeoff between dipole size and number.
    The analyzed models include dipole lengths from $0.01$ to $5$ nm, with portions of this range where the probability was zero have been trimmed.
    A peak is visible along this tradeoff curve, showing the most likely combination of TLS density and lengths.  The ridge in the plot shows a rough $\ell \propto n_{T}^{-1/2}$, as would be expected from Eq.~\ref{eq:vv}. The key results are the peak around $\ell = 0.3$ nm and $n_T = 6 \times 10^{10}$/cm$^{2}$, and the fairly sharp dropoff away from the peak.}
    \label{fig:connorsplot}
\end{figure}

The results of the data analysis are shown in Fig.~\ref{fig:connorsplot}.
The configurations under test show the expected trade-off between dipole quantity and magnitude, with smaller dipoles needing to appear in greater numbers to agree with the measured noise.  The key result is an expected value for the count $\overline{N}_T$ according to the formula
\begin{equation}
    \overline{N}_T = \sum N_T P(\mathcal{M}(N_T, p, i_{\mathrm{or}})),
    \label{eq:expectationvaluedefinition}
\end{equation}
where the sum ranges over all possible values of the $N_T$, $p$ and $i_\mathrm{or}$.
The expected value for the count is $63.18$, which corresponds to an expected TLS density of $\SI{9.02e10}{\cm^{-2}}$.
This compares well with previously published estimates on other devices \cite{shehata2023modeling, kkepa2023simulation, kkepa2023correlations}.
The expected value of the dipole moment, calculated analogously to the expected count, is $p_0 = e \ell$ where $\ell = \SI{0.186}{\nm}$.
This matches a result obtained by \textcite{hung2022probing} for the largest dipole moments they obtained in amorphous alumina (though this is a rather different system).

As mentioned above, the models considered here assume that all noise dipoles have the same magnitude.  Thus the possibility of a very broad range of magnitudes is left open.

By restricting the models used in the expectation value, we calculate conditional expectation values assuming particular conditions are true.
Table~\ref{tab:connorsExpectationValuesOrientation} shows such conditional expectation values by orientation.
Our results for the noise measured in Ref.~\cite{connors2022charge} indicate that both vertically-and horizontally-oriented dipoles are possible, but vertical dipoles would need to be smaller and more numerous than horizontal ones.

\begin{table}
    \begin{tabular}{c | c c c | c}
       Dipole & Vertical & Horizontal & Random & Combined \\
        \hline
    Length ($\si{\nm}$) & $0.135$ & $0.288$ & $0.188$ & $0.186$  \\
    Density ($\SI[print-unity-mantissa = false]{1e10}{\cm^{-2}}$) & 10.73 & 5.94 & 9.54 & 9.03 \\
    \end{tabular}
    \caption{Expectation values for the predicted dipole length and dipole number density deduced from the data of \textcite{connors2022charge}, conditioned on a particular orientation.
    The combined column represents an unconditioned expectation value, as defined in Eq.~\ref{eq:expectationvaluedefinition}. It is the average over all three orientations, weighted by the probability calculated for that orientation.
    The other results are the expectation values calculated conditioned on each particular orientation.
    For instance, if we are told that the noise dipoles are vertically oriented, then the expectation value for dipole length is $\SI{0.135}{\nm}$ and the number density is $\SI{10.73e10}{\cm^{-2}}$.}
    \label{tab:connorsExpectationValuesOrientation}
\end{table}

\subsection{Analysis of CPSD}
\label{subsec:bayescpsd}

We extend our analysis from APSD to CPSD with the goal of seeing how much additional information about the TLS can be obtained by using two qubits instead of one.  The data are those of  \textcite{yoneda2023noise}.
They measure fluctuations in qubit precession rates for each of a pair of spin-qubits $\qty{100}{\nm}$ apart in a Si/SiGe double quantum dot.
The qubit precession rate is proportional to electric field in the $x$ direction, meaning they measure fluctuations in the electric field.
This differs from the noise measurements from \textcite{connors2022charge} analyzed above, where instead electric potential is measured.
Thus we need Eq.~\ref{eq:ee} instead of Eq.~\ref{eq:vv}.
We choose a rectangular region $\SI{400}{\nm} \times \SI{200}{\nm}$ ranging from $0 < z < \SI{50}{\nm}$.
This range for $z$ is different from the oxide layer thickness used in \textcite{connors2022charge}, where there is an assumed spatial gap in the $z$ direction below the bottom face of the oxide layer.

As in the earlier analysis, we sweep across a variety of TLS counts and dipole moment magnitudes, calculating the likelihood for each combination of parameter values.
This is used to find the probability of a set of parameters $P(\mathcal{M}(N_T, p, i_{\mathrm{or}}))$, as with the earlier results.
The two quantum dots measured by \citeauthor{yoneda2023noise} are assumed to be positioned at $\left(x, y, z\right) = \left(\pm \SI{50}{\nm}, 0, 0\right)$.

We calculate the noise likelihood estimates with four different simulated measurements:
the APSD for dot 1 only, the APSD for dot 2 only, the APSDs of both  dot 1 and dot 2, and finally doing the full analysis that combines the APSDs of both dots 1 and 2 together with the phase of the CPSD between dot 1 and dot 2.
For each of the two dots, we compare the likelihood of obtaining a matching sample based on the individual APSDs along with the combined data.
This can provide a sense of how much additional information the CPSD phase measurement provides, by comparing the likelihoods obtained with and without the CPSD measurements.

\begin{figure*}[ht]
\centering
\setkeys{Gin}{width=\linewidth}
	\begin{subcaptionblock}{.4\textwidth}
		\includegraphics{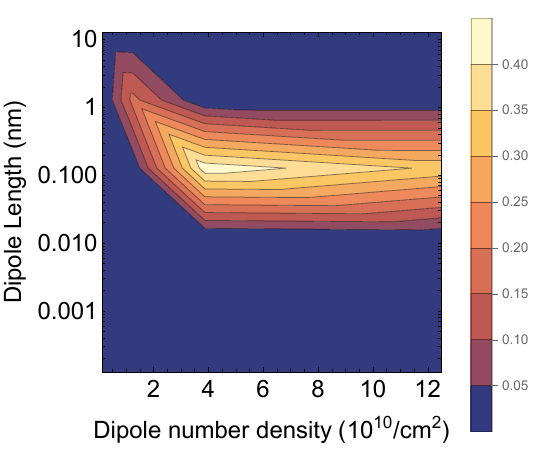}
		\caption{Only APSD data}
		\label{fig:taruchaAll:ApsdOnly}
	\end{subcaptionblock}%
	\begin{subcaptionblock}{.4\textwidth}
		\includegraphics{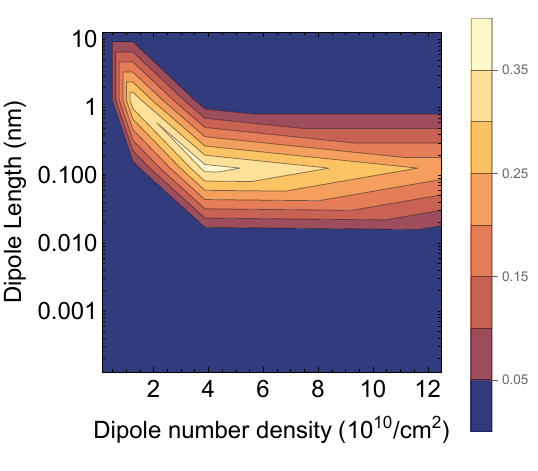}
		\caption{APSD and CPSD phase}
		\label{fig:taruchaAll:ApsdPhase}
	\end{subcaptionblock}%
	\caption{
	Probability assigned to parameters from an analysis of the data from \textcite{yoneda2023noise}.
	The probabilities are summed over different orientation types.
	Results are presented for (a) the analysis using only APSD data and (b) additionally filtering using the CPSD phase data.
	Including the phase data pushes the probabilities towards fewer, larger TLSs.
	\label{fig:taruchaAll}
	}
\end{figure*}

The calculated probability distributions $P(\mathcal{M}(n_T, l))$ for different parameter values are plotted in Fig.~\ref{fig:taruchaAll}.
Fig.~\ref{fig:taruchaAll}(a) contains the results for the analysis using APSD data only.
The expectation value of the number of TLSs used was 50.499, which over a $\qty{400}{\nm} \times \qty{200}{\nm}$ area corresponds to a TLS density of $\qty{6.31e10}{\cm^{-2}}$.
This is the same order of magnitude as the results presented in Sec.~ \ref{subsec:bayesapsd}.
Similarly, the dipole length $\ell$ is estimated from the dipole moment magnitude by assuming the charge is that of a single electron: $p=|e| \ell$.
For the analysis using only the APSD data at the two dots, the expectation value of $\ell$ is \qty{0.196}{\nm}.

We should also note that the conversion of electric field noise to qubit frequency noise requires a proportionality constant estimated by \citeauthor{yoneda2023noise}. This introduces an error of order unity.
Nevertheless, the numerical value is similar to that obtained in \cref{subsec:bayesapsd}.
This is true even though the sweep over dipole moment magnitude ranged over five orders of magnitude.
The approximate agreement of both density and dipole magnitude is encouraging since it suggests a measure of universality across devices.

We now add the information given by the phase data of the CPSD. Unlike the noise data, phase data is angular and regression and other techniques require special handling \cite{Mardia1999}.
Furthermore, the only possible simulated values are $\widetilde{\gamma} = 0$ or $\pi$.  Gradient methods do not apply.
In other words, the phase data is either correct or incorrect, and does not give any information on the direction parameters should change to minimize cost.
This means that it is less useful as a cost function, in terms of quantifying the degree of difference in PSD.
However, it remains useful as a filter in the Monte Carlo process, where it can rule out incorrect configurations that are otherwise not ruled out.

\begin{figure*}[ht]
\centering
\setkeys{Gin}{width=\linewidth}
	\begin{subcaptionblock}[b]{.48\textwidth}
		\includegraphics[width=\linewidth]{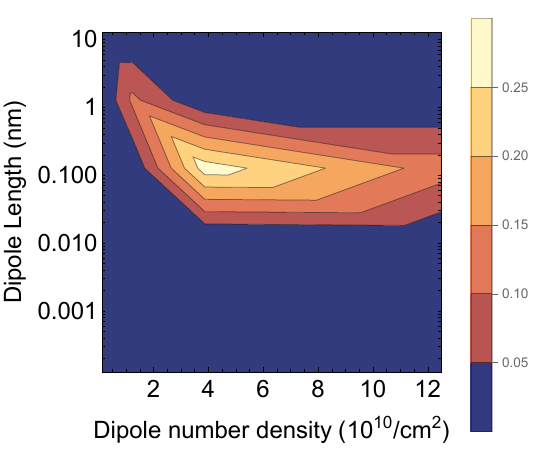}
		\caption{
			APSD-only, $XY$-oriented dipoles.
		}
		\label{fig:taruchaApsdOnlyXY}
	\end{subcaptionblock}%
	\begin{subcaptionblock}[b]{.48\textwidth}
		\includegraphics[width=\linewidth]{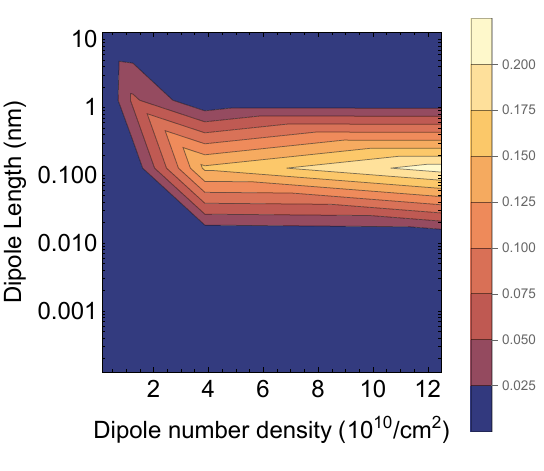}
		\caption{
			APSD-only, $Z$-oriented dipoles.
		}
		\label{fig:taruchaApsdOnlyZ}
	\end{subcaptionblock}\\
	\begin{subcaptionblock}[b]{.48\textwidth}
		\includegraphics[width=\linewidth]{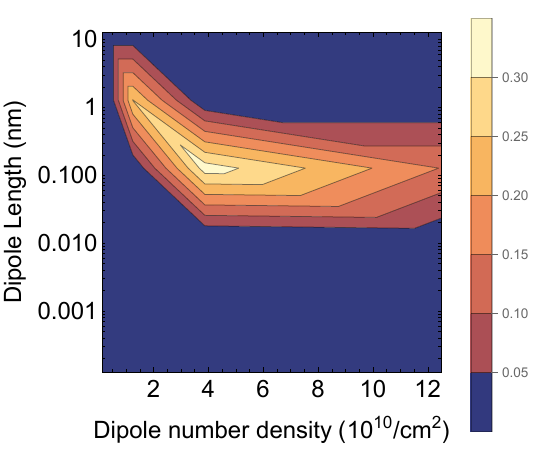}
		\caption{
			APSD+Phase, $XY$-oriented dipoles.
		}
		\label{fig:taruchaApsdPhaseXY}
	\end{subcaptionblock}%
	\begin{subcaptionblock}[b]{.48\textwidth}
		\includegraphics[width=\linewidth]{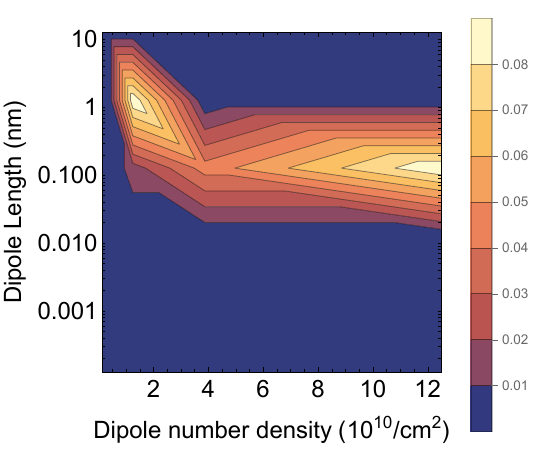}
		\caption{
			APSD+Phase, $Z$-oriented dipoles.
		}
		\label{fig:taruchaApsdPhaseZ}
	\end{subcaptionblock}%
	\caption{
		Comparison of dipole orientation effects on parameter probability distributions from analysis of the \textcite{yoneda2023noise} data.
		Vertical dipoles show a tendency for a larger number of smaller dipoles than horizontal.
		Top row (a,b) shows results using only APSD data, while bottom row (c,d) includes both APSD and CPSD phase data.
		Left column (a,c) shows horizontal ($XY$) dipoles, while right column (b,d) shows vertical ($Z$) dipoles.
		The phase data has a much stronger effect on constraining the vertical dipole parameter space than the horizontal one.
		This suggests that the CPSD phase measurements are potentially valuable for distinguishing between different dipole orientations.
	}
	\label{fig:tarucha_orientation_comparison}
\end{figure*}

Including the phase data into the cost function and performing the same Bayesian analysis as with the APSDs only, we obtain the probability plots shown in Fig.~\ref{fig:taruchaAll} (b) and broken down by orientation in Fig.~\ref{fig:tarucha_orientation_comparison}.
The inclusion of the CPSD phase data increases probability in the direction of fewer, larger dipoles.
This shows that the phase data does provide information not contained within the APSDs alone.
\Cref{fig:tarucha_orientation_comparison} shows that the primary impact of the inclusion of phase data is to change the probability results for vertically oriented dipoles.

Overall, these measurements by \textcite{yoneda2023noise} and \textcite{connors2022charge} give similar results for the dipole densities and magnitudes.
The phase data makes a difference in distinguishing between orientations, which may be difficult to do using only APSD measurements.

\section{Conclusion}
\label{sec:conclusion}

The broadly accepted model of fluctuating two-level electric dipoles creating noise on qubits was investigated. The aim is to deduce the positions and other physical characteristics of the dipoles using qubits as noise sensors that give the APSD and the CPSD.  Several experimental studies of this type have appeared in the literature.

We gave evidence that the problem is underdetermined - the solution space is too large given the amount of information we can hope to get from experiments. This evidence came in the form of pointing out a mismatch in the number of parameters to be determined and the relative paucity of data, and in an example that showed the ineffectiveness of standard curve-fitting.
This implies that in nearly all cases, it is hopeless to try to determine all of the physical parameters of the TLS.

However, useful partial information may still be obtainable. We propose two information categories. 1. Qualitative patterns of the TLS configuration can be obtained from comparison with analytic calculations. 2. Probabilistic information about the configuration of the TLS can be inferred from Bayesian analysis.   
We studied the phase and correlation strength of CPSD for the electric potential at quantum dots. This led to a geometric interpretation for the phase of CPSD. The strongest conclusions can be drawn when the noise is anti-correlated on two dots.  This tells us that the TLS are relatively close to the dots and that the dipoles are oriented along the separation vector of the dots.  This can be useful information for designing experiments to investigate the physical nature of the TLS and their positions and the magnitude of the dipoles. The additional information in the CPSD suggests that multiple-dot setups could be very useful. A recent experiment of this type measures quantum dots that are located at different heights ($z$ coordinates)~\cite{ivlev2024coupled}, which would give information complementary to that which currently exists.  

Probabilistic information is obtained by applying a Bayesian method to the experimental data of both APSD and CPSD measurements. This method can give information on all the parameters of the TLS, but in this paper we focused mainly on the number density and size of TLS that can fit the data in Ref.~\cite{connors2022charge}  and Ref.~\cite{yoneda2023noise}. This provides estimates of Bayesian likelihoods over a sweep of parameter values for number density and TLS size, as well as providing expectation value estimates for both parameters. The results showed that the data measured in \cite{yoneda2023noise} is most compatible with vertically oriented TLS. By evaluating the Brier score, the quality of information coming from a noise measurement was assessed in a quantitative fashion.

\section{Acknowledgments}
We thank Leah Tom for useful discussions. This research was sponsored by the Army Research Office (ARO) under Awards No.\ W911NF-17-1-0274 and No.\ W911NF-23-1-0110. YC acknowledges support from the Army Research Office through Grant No.\ W911NF-23-1-0115.
The views, conclusions, and recommendations contained in this document are those of the authors and are not necessarily endorsed nor should they be interpreted as representing the official policies, either expressed or implied, of the Army Research Office (ARO) or the U.S.\ Government. The U.S.\ Government is authorized to reproduce and distribute reprints for Government purposes notwithstanding any copyright notation herein.
We also thank the University of Wisconsin-Madison Center for High-Throughput Computing.

\bibliography{main}

\end{document}